\documentclass[12pt]{article}
\usepackage{cite,epsfig,amssymb,amsmath,graphicx,color,fleqn}

\newcommand{\RNum}[1]{\uppercase\expandafter{\romannumeral #1\relax}}

\newcommand{\be}{\begin{equation}}
\newcommand{\ee}{\end{equation}}
\newcommand{\bear}{\begin{eqnarray}}
\newcommand{\eear}{\end{eqnarray}}
\newcommand{\ba}{\begin{array}}
\newcommand{\ea}{\end{array}}
\newcommand{\nn}{\nonumber}

\usepackage[normalem]{ulem}
\usepackage{varioref}

\begin{document}

\begin{center}
{{{\Large \bf Massive Photon and Dark Energy }
}\\[17mm]

Seyen Kouwn$^{1}$,~~Phillial Oh$^{2}$, and~~Chan-Gyung Park$^{3}$\\[3mm]

{\it$^{1}$Korea Astronomy and Space Science Institute,
Daejeon 305-348, Republic of Korea\\[2mm]
$^{2}$Department of Physics,~BK21 Physics Research Division,
~Institute of Basic Science,\\
Sungkyunkwan University, Suwon 440-746, Korea\\[2mm]
$^{3}$Division of Science Education and Institute of Fusion Science,
Chonbuk National University, Jeonju 561-756, Korea}\\[2mm]
{\tt seyenkouwn@kasi.re.kr,~ploh@skku.edu,~parkc@jbnu.ac.kr} }
\end{center}

\vspace{10mm}

\begin{abstract}
We investigate cosmology of massive electrodynamics and
explore the possibility whether massive photon could provide
an explanation of the dark energy.
The action is given by the scalar-vector-tensor  theory of gravity which is obtained by non-minimal coupling of the massive Stueckelberg QED with gravity and
its cosmological consequences are studied by paying a particular attention to the role of photon mass. We find that the theory
allows cosmological evolution where the radiation- and  matter-dominated epochs are followed by a long period of virtually constant dark energy that closely mimics $\Lambda$CDM model and the main source of the current acceleration is provided by the nonvanishing photon mass governed by the relation
$\Lambda\sim m^2$.
A detailed numerical analysis shows that the nonvanishing photon mass of the order of $\sim 10^{-34}$ eV is consistent with the current observations. This magnitude is far less than the most stringent limit on the photon mass available so far, which is of the order of $m \leq  10^{-27}$eV.
\end{abstract}
\newpage

\tableofcontents

\newpage
\section{Introduction}
The intriguing discovery of the current accelerating Universe \cite{Riess:1998cb,Perlmutter:1998np} has generated extensive investigations searching for the foundation which provides a theoretical explanations.
The simplest $\Lambda$CDM model \cite{Komatsu:2008hk} with the cosmological constant $\Lambda$ as dark energy fits very well
with observations and is regarded as the most accepted approach.
To be consistent with current observations, the cosmological constant $\Lambda$ has to be a very small number in value: $\sim 10^{-120}$ orders of magnitude smaller than the Planck scale $M_{\rm p}$ and this extreme fine-tuning leads to the cosmological constant problem \cite{Weinberg:1988cp}. It has invoked other diverse attempts in search of the origin of dark energy. One of the alternative methods is the quintessence model \cite{Ratra:1987rm} in which the  cosmological constant is a dynamically varying potential energy of a scalar field. Despite of its simplicity and many attractive features, some problems remain. For example, it does not resolve the puzzle of cosmic coincidence, which seems to be a common feature of decaying cosmological constant with a few exceptions \cite{Bludman:2004az}. It also necessitates the introduction of a new scalar field, whereas the only experimentally verified available scalar field is the Higgs field.

One of the possible pathways to the cosmological constant problem may be to suppose that $\Lambda$ is associated with other  fundamental mass scales;
Possible candidates are the UV cutoff of quantum field theory based on holographic principle \cite{Cohen:1998zx}, or the electroweak scale
\cite{ArkaniHamed:2000tc}, which can provide a natural explanation of the coincidence problem. Another possibility is that its smallness is related with yet another  small number of nature. Then, it is conceivable that there might exist some relation connecting them. It could come as a solution of the equations of motion, or might be a consequence of fundamental reason which is inaccessible now.  The first candidate that comes to remembrance is the mass of the photon, if it has a mass at all.
In this paper, we investigate cosmology of massive electrodynamics and
explore the possibility whether massive photon could provide
an explanation of the dark energy.




The photon mass is usually assumed to be exactly
zero. This is based on the Maxwell equations which
describes massless photon.
In addition, a photon mass term in quantum electrodynamics breaks
gauge invariance and might spoil the renormalizability, which renders the theory
quantum mechanically inconsistent.
However, the consideration of non-vanishing photon mass  \cite{Tu:2005ge}
has a long history, and theoretically, it is well known  that Maxwell theory with abelian gauge symmetry
can be extended to a gauge invariant massive theory  by
means of the Stueckelberg mechanism
\cite{Ruegg:2003ps}{\footnote{The photon  can also become massive
 through spontaneous symmetry breaking via Higgs mechanism
  \cite{Suzuki:1988bd}, but this idea will not be pursued in this paper}}.
It introduces a scalar field which compensates the gauge transformation of the vector field.
Such massive theory preserves the unitarity and renormalizability
of the massless theory. Moreover, the possible conflict between the massive QED and standard model could be avoided \cite{Heeck:2013cfa}.
In a particular gauge where the scalar field is set to zero, the massive theory reduces to the Proca theory which describes
electrodynamics of massive vector field \cite{Accioly:2010zzb}.
The question of a photon mass in QED should then be tested
 experimentally.
If there is any deviation from zero, it must be very
small, because Maxwell theory has been verified
to an extreme accuracy. On the other hand, the experimental constraints on the photon mass has considerably increased over the past several
decades, putting upper bounds on its mass. So far, the most stringent upper limit is given by $m \leq 10^{-27}{\rm eV}$ \cite{Hagiwara:2002fs}. In all these researches, the photon is described by massive  Proca theory, which does not include the Stueckelberg
field.




There exist many attempts to link the cosmology of vector fields
with accelerating
Universe \cite{Ford:1989me,Jimenez:2008au}, but a direct cosmological consequences of the massive QED in relation with the dark energy  has been considered only recently \cite{Kim:2013wsa}.
It was shown that the massive QED without the Stueckelberg field
(but with the nonvanishing torsion components) has the potential of possible explanation of the dark energy in terms of the photon mass, where the dark energy density (cosmological constant), which is proportional to the photon mass squared, is allowed as a solution of equations of motion.
In this work, we take the full massive QED including the scalar field and investigate the cosmology. The theory consists of massive vector field and Stuckelberg scalar field
interacting with Einstein gravity. Also, for general purposes, we include
non-minimal interaction
terms in which the vector field interacts with scalar curvature and Ricci tensor.

The action contains a scalar field which is necessary to endow the photon with mass while preserving the gauge invariance. Possible cosmological consequences of this Stueckelberg field were considered before\cite{Jimenez:2012ak}. Its role in  ordinary massive QED is to cancel the contribution of the unphysical pole of the vector propagator in the physical processes, and it cannot appear as physical states\cite{Aitchison:1980mv}
.
This is evident because the field can be completely gauged away in the unitary gauge. However, such decoupling of the Stueckelberg field does not operate when the gravitational interaction is included.
The gravitational coupling can accommodate non-vanishing contributions of the Stueckelberg field. For example, in the massless limit of massive Proca theory,
the longitudinal scalar mode remains coupled to gravitation, even though it is decoupled from the current \cite{Deser:1972wi}
. The same reasoning will apply to the Stueckelberg
scalar field with the covariant massive QED: It will also decouple from the current in the massless limit, but the gravitational coupling remains. Therefore,  it cannot be neglected in cosmology, and effectively we are considering non-minimally interacting scalar-vector-tensor  theory of gravity.

It is worth mentioning the gauge invariance of energy-momentum tensor of the covariant action.  If we calculate the energy-momentum tensor of the covariant action of the Proca theory, for example,   it will contain a gauge dependent piece coming from the gauge-fixing term. However, this term becomes null, if we apply the Lorentz gauge condition
(See Eq. (\ref{gaugefixing})), and the gauge invariance of the energy-momentum tensor is intact in quantum field theory. But as far as cosmology is concerned, we might accept the effective action as a classical one and  attempt to look for time-dependent behaviour of the gauge field with only temporal component being non-vanishing. This is necessary in order to  respect the isotropy and homogeneity. In general, this will bring in a gauge dependence of the energy-momentum tensor\cite{Jimenez:2008au}, but this should not imply the inconsistency of the cosmological approach, but could be taken
as an indication of a characteristic of the gravitational interaction.
It is interesting to note that in the pure electromagnetic case, the
gauge fixing term with only the temporal component of the gauge potential induces a  vacuum energy or a cosmological constant whose value depends on the gauge-fixing parameter, but this is still harmless to the ordinary QED.


The purpose of this work is to study the cosmology of the scalar-vector-tensor (SVT) (or Einstein-Proca-Stueckelberg) theory of gravity  which is obtained by non-minimally coupling massive QED with gravity and compare the results with the observations, especially focusing on the photon mass.
The SVT theory has several parameters whose number is to be restricted by the
observational constraints. A  couple of the parameters are related with the cosmological solution
which yields both decaying and growing modes and they can be fixed
from the beginning by choosing the decaying mode conditions. These conditions
allow cosmological evolution in which the radiation- and  matter-dominated epochs are entailed by a long period of virtually constant dark energy, which mimics $\Lambda$CDM. The main source of the dark energy is provided by the nonvanishing photon mass during this period. A detailed numerical analysis shows that the nonvanishing photon mass of the order of $\sim 10^{-34}$ eV is consistent with the current observations. This magnitude is far less than the most stringent limit on the photon mass available so far, which is of the order
of $m \leq  10^{-27}$eV \cite{Hagiwara:2002fs}.






The paper is organized as follows: In Sec.~\ref{secmodel}, we construct Einstein-Proca-Stueckelberg theory of massive QED interacting with gravity and write down equations of motion for FRW cosmology.
In Sec.~\ref{secevol}, we analyze the cosmological evolutions in the radiation, matter, and dark energy dominated epochs, respectively.
In Sec.~\ref{secob}, observational constraints on our model parameters are presented. Sec.~\ref{seccon} includes conclusion and discussions.

\section{Model}\label{secmodel}
The  action we consider is the gauge fixed massive QED theory which is non-minimally interacting with the Einstein gravity:
Keeping  terms only up to second derivative of the fields brings to the  following
\begin{align}
S = \int d^4x \sqrt{-g}\Biggr[ & {R\over 2 \kappa} -{1\over4}F_{\mu\nu}F^{\mu\nu}-{1\over 2} m^2 A_\mu A^\mu
-{1\over2\xi}\left(\nabla_\mu A^\mu\right)^2
-{1\over2}\nabla_\mu \phi \nabla^\mu \phi -{1\over 2}\xi m^2 \phi^2
\nonumber  \\
&+\omega A_\mu A^\mu R+ \eta A^\mu A^\nu R_{\mu\nu}
+ {\chi\over 2}\phi^2  R
\Biggr]\,, \label{qedaction}
\end{align}
where $F_{\mu\nu} = \partial_\mu A_\nu - \partial_\nu A_\mu$,  $\xi$ is  the gauge
fixing parameter,
and $\omega$, $\eta$ and $\chi$ are dimensionless parameters describing the non-minimal interactions.

A couple of comments are in order. The above action (\ref{qedaction}) in flat space reduces to the massive QED with Stueckelberg scalar field in the covariant gauge. It is the most general second derivative action which describes the non-minimal interaction of Stueckelberg scalar and massive vector field with the Einstein gravity, and  belongs to the most simple scalar extension of the vector-metric theory of
gravity \cite{Hellings:1973zz}.


The Einstein equations obtained from action \eqref{qedaction} by
varying with respect to the metric $g_{\mu\nu}$ can be written in
the following way:
\begin{align}
{1\over\kappa}G_{\mu\nu}
= T^{(\varphi)}_{\mu\nu}+m^2T^{(m^2)}+T^{(F_{\mu\nu})}_{\mu\nu}
-{1\over 2 \xi}T^{(\xi)}_{\mu\nu}+\omega \,T^{(\omega)}_{\mu\nu}+\eta\, T^{(\eta)}_{\mu\nu}
+{\chi\over2}\,T^{(\chi)}_{\mu\nu}
+T^{(m,r)}_{\mu\nu}\,,
\end{align}
where $T^{(m,r)}_{\mu\nu}$ is the energy-momentum tensor corresponding
to other fields (matter and radiation) and we have defined
\begin{align}
&T^{(\varphi)}_{\mu\nu}
= \nabla_\mu \varphi \nabla_\nu \varphi+g_{\mu\nu}\left(
-{1\over2}\nabla_\alpha \varphi \nabla^\alpha \varphi -V(\varphi)
\right) \,, \\
&T^{(m^2)}_{\mu\nu}
= A_\mu A_\nu +g_{\mu\nu}\left(
-{1\over2}A_\alpha A^\alpha
\right) \,, \\
&T^{(F_{\mu\nu})}_{\mu\nu}
= F_{\mu}^{~\alpha}F_{\nu\alpha}+g_{\mu\nu}\left(
-{1\over4}F_{\alpha\beta}F^{\alpha\beta}
\right) \,, \label{maxzz}\\
&T^{(\xi)}_{\mu\nu}
= 4 A_{(\mu} \nabla_{\nu)} \nabla_\alpha A^\alpha-g_{\mu\nu}\left(
\left(\nabla_\alpha A^\alpha\right)^2
+2A^{\alpha}\nabla_\alpha\nabla_\beta A^\beta
\right) \,, \label{gaugefixing} \\
&T^{(\omega)}_{\mu\nu}
= 2\left(
\nabla_{(\mu}\nabla_{\nu)}A^2 -A_\mu A_\nu R
-A_\alpha A^\alpha G_{\mu\nu}-g_{\mu\nu} \square A^2
\right) \,, \label{omegamax}\\
&T^{(\eta)}_{\mu\nu}
= 2\nabla_{\alpha}\nabla_{(\mu}A_{\nu)}A^{\alpha}
-4A^{\alpha} R_{\alpha(\mu}A_{\nu)}
-\square A_\mu A_\nu \nn \\
&~~~~~~~~~~~~~~~~~~~~~~~+g_{\mu\nu}\left(
 A^\alpha A^\beta R_{\alpha\beta}
- \nabla_\alpha \nabla_\beta A^\alpha A^\beta
\right) \,,\label{etamax} \\
&T^{(\chi)}_{\mu\nu}
= 4\nabla_\mu \varphi \nabla_\nu \varphi
-2 \varphi^2 G_{\mu\nu}+4 \varphi \nabla_{(\mu}\nabla_{\nu)}\varphi
-2g_{\mu\nu} \square \varphi^2 \,,
\end{align}
where $\Box=\nabla_\mu \nabla^\mu$,  $A^2 = A_\mu A^\mu$
and brackets in a pair of
indices denoting symmetrization with respect to the corresponding
indices.
Apart from the Einstein equations we can obtain a
set of field equations for gauge $A_\mu$ and scalar fields $\varphi$
by varying the action with respect to the vector and scalar field to give
\begin{align}
&\nabla_\nu F^{\mu\nu}+\left(m^2-2\omega R\right) A^\mu
- 2\eta R^\mu_{~\nu}A^\nu
-{1\over \xi}\nabla^\mu\left(\nabla_\alpha A^\alpha\right)
= 0 \,, \\
&\square\varphi - \left(\xi m^2 - \chi R\right)\varphi= 0
\,.
\end{align}

In this work we shall study the isotropic and
homogeneous flat cosmology. Thus, we consider   the 
time dependent  vector field and scalar field, 
so that{\footnote{Note that the configuration (\ref{vectorc}) gives $F_{\mu\nu}=0$, and does not contribute to the photon radiation energy.
Also, we assume that the {\it spatial} average of the photon polarization vector $\vec A$ is zero and mixing between $A_0$ and $\vec A$ in Eqs. (\ref{gaugefixing})-(\ref{etamax})
can be neglected. The contribution of quadratic terms in $\vec A$ 
is treated separately  and is included as the photon radiation energy in $T^{(r)}_{\mu\nu}$. }  }
\begin{align}
A_{\mu} = \Big(f(t),0,0,0\Big) \,, \quad
\varphi = \varphi(t)
\,,\label{vectorc}
\end{align}
and the space-time geometry
is  given by the flat Robertson-Walker metric:
\begin{align}
ds^2 = -dt^2 + a^2(dx^2+dy^2+dz^2)
\,.
\end{align}
In this metric, the field equations for the vector and scalar can be rewritten as
\begin{align}\label{eqgauge}
&\ddot{f}+3 H \dot{f}+3 \dot{H}f
+\xi f \left[ m^2-6\left(\eta+4\omega\right) H^2
-6\left(\eta+2\omega\right)\dot{H}\right] = 0
\,,\\
&\ddot{\varphi}+3H\dot{\varphi}-6\chi \left(2H^2+\dot{H}\right) \varphi+ m^2 \xi \varphi = 0
\,,
\end{align}
and Einstein equations as follows;
\begin{align}\label{eqH}
{3\over \kappa}H^2=&
\rho^{(r)}+\rho^{(m)}+\rho^{{\rm (de)}}
\,, \\
-{3\over \kappa}H^2-{2\over \kappa}\dot{H}=&
 p^{(r)} +p^{(m)} + p^{{\rm (de)}}
\,,
\end{align}
where $H\equiv \dot{a}/a$ is the Hubble parameter and we added the standard radiation and matter energy densities. $\rho^{{\rm (de)}}$ and $p^{{\rm (de)}}$ are the energy density and pressure coming from the temporal component of  the vector (tv) plus  scalar (s) fields and interpret   $\rho^{{\rm (de)}}$ and $p^{{\rm (de)}}$ as  the dark energy density and dark pressure.
They are  respectively given  as follows
\begin{align}
\rho^{{\rm (de)}} = \rho^{{\rm (tv)}}  + \rho^{{\rm (s)}}\,, \quad
p^{{\rm (de)}} = p^{{\rm (tv)}}  + p^{{\rm (s)}} \,, \label{energypressure}
\end{align}
where
\begin{align}
\rho^{{\rm (tv)}} \equiv&
{1\over \xi}f\ddot{f}-{1\over 2\xi}\dot{f}^2+{1\over 2}m^2 f^2
+6\left(\eta+2\omega\right)Hf\dot{f}
-\left({9\over 2\xi}+18\omega\right)H^2f^2
-\left(6\eta-{3\over \xi}+12\omega\right)\dot{H}f^2 \,, \\
 \rho^{(\rm s)}\equiv&
{1\over2}\dot{\varphi}^2+{1\over2}m^2\xi \varphi^2
-6\chi H\varphi\dot{\varphi}
-3\chi H^2\varphi^2 \,, \\ 
p^{{\rm (tv)}} \equiv &
\left(-2\eta+{1\over \xi}-4\omega\right)f\ddot{f}+\left(-2\eta+{1\over 2\xi}-4\omega\right)\dot{f}^2
+{1\over2}m^2 f^2  
+\left(-8\eta+{6\over \xi}-8\omega\right)H f\dot{f} \,, \nonumber \\
&+\left(-6\eta+{9\over 2\xi}-6\omega\right)H^2f^2 
+\left(-4\eta+{3\over \xi}-4\omega\right)\dot{H}f^2  \\
p^{{\rm (s)}} \equiv &
2\chi \varphi \ddot{\varphi} +\left({1\over2}+2\chi\right)\dot{\varphi}^2
- {1\over2}m^2\xi \varphi^2
+ 4\chi H \varphi\dot{\varphi}
+ 3\chi H^2 \varphi^2 
+ 2\chi \dot{H} \varphi^2
\,.\label{eqHpress}
\end{align}


\section{Cosmological Evolution}\label{secevol}
In this section we analyze the evolution equations by assuming that the universe in each stage is dominated by a barotropic perfect fluid with constant equation of state parameter $w_i=\rho_i/p_i~(i=m,r)$
and later   by $\rho^{{\rm (de)}}$. Then, we check our results numerically.

\subsection{Radiation dominated epoch}
In  the radiation dominated epoch, let us assume that the energy densities of the matter and the dark energy are negligible,
\begin{align}\label{radcondi}
\rho^{\rm (de)} \ll \rho^{(r)} \,, \quad
\rho^{(m)} \ll \rho^{(r)} \,,
\end{align}
so that the Hubble parameter $H$ is given by
\begin{align}
{3\over \kappa}H^2 \simeq  {\rho_{r,0} \over a^{4}} \,,
\end{align}
where $\rho_{r,0}$ is the present value of the energy density of the radiation.
In such a case, we obtain the field equations, which can be written as
\begin{align}
&a^2 f'' + 2af' + 6 \left(\eta  \xi-1\right) \simeq 0 \,, \\
&a^2 \varphi'' + 2a\varphi' \simeq 0,
\end{align}
where we neglect the mass of the photon with respect to the Hubble
parameter and prime is a derivative with respect to the scale factor.
They have the following solutions
\begin{align}
f(a) &=  c^{(r)}_{-} a^{-\frac{1}{2}-\frac{\sqrt{25- 24\eta \xi}}{2}}
+ c^{(r)}_{+} a^{-\frac{1}{2}+\frac{\sqrt{25-24\eta \xi}}{2}}
\,, \label{radsolf} \\
\varphi(a) &={ d^{(r)}_{-} \over a}+ d^{(r)}_{+}  \,, \label{radsolphi}
\end{align}
where $c^{(r)}_\pm$ and $d^{(r)}_\pm$ are integration constants.
Here,  we impose the conditions, $c^{(r)}_-=0$ and $d^{(r)}_-=0$
in order to  make the solutions non-singular as $a\rightarrow 0$.
Inserting the above solutions into \eqref{energypressure}
and assuming that $c^{(r)}_{+}$ and $d^{(r)}_{+}$ are being of order ${\cal O}(1)$, 
we obtain that the conditions \eqref{radcondi} yields restrictions
\begin{align}
\eta \xi \lesssim 1 \,,\quad
\chi \lesssim 0 \,.
\end{align}

Concerning the evolutions of the temporal component of the gauge field and scalar field,
according to \eqref{energypressure},
the dark energy and pressure are given by
\begin{align}
\rho^{\rm (de)} \simeq -\chi{(d^{(r)}_+)^2 \kappa \rho_{r,0} \over a^{4} }
\,,\quad
p^{\rm (de)} \simeq -{\chi\over3}{(d^{(r)}_+)^2 \kappa   \rho_{r,0}  \over a^{4} }
  \,.\label{RD}
\end{align}
Here, we find that $\rho^{\rm (tv)}$ and $p^{\rm (tv)}$ are much smaller than those of scalar field and have neglected them.
This is because the temporal component is proportional to the scale factor in such a way that its energy density and pressure scale with an exponent larger than $-4$.
Thus the leading behavior of the energy density and pressure  comes from the scalar field.
We can also calculate the equation of state parameter which results in
\begin{align}
w^{\rm (de)} \equiv {p^{\rm (de)}\over \rho^{\rm (de)}} \simeq {1\over 3}\,.
\end{align}
Here, we note that the dark energy scales as radiation.
Therefore, during the radiation epoch the fraction of energy density, $\rho^{(r)}/\rho^{\rm (de)}$, is a constant.

\subsection{Matter dominated epoch}
In  the matter dominated epoch,
we assume that the energy densities of the radiation and
 the dark energy are negligible,
\begin{align}\label{matcondi}
\rho^{\rm (de)} \ll \rho^{(m)} \,, \quad
\rho^{(r)} \ll \rho^{(m)} \,,
\end{align}
so that the Hubble parameter $H$ is given by
\begin{align}
{3\over \kappa}H^2 \simeq  {\rho_{m,0} \over a^{3}} \,,
\end{align}
where $\rho_{m,0}$ is the present value of the energy density of the matter.
In such a case,  the evolution equations can be written as
\begin{align}
& a^2 f'' + \frac{5}{2} a f' + \left(3 \eta  \xi -6 \xi  \omega -\frac{9}{2}\right) f \simeq 0 \,, \\
& a^2 \varphi ''(a)+\frac{5}{2} a \varphi '(a) \simeq 0,
\end{align}
and they have the following solutions
\begin{align}
&f(a) \simeq c^{(m)}_- a^{-\frac{\beta }{4}-\frac{3}{4}}
+c^{(m)}_+ a^{\frac{\beta }{4}-\frac{3}{4}}  \,,\label{decaying} \\
&\varphi(a) \simeq \frac{d^{(m)}_-}{a^{3/2}}+d^{(m)}_+  \,.\label{SD}
\end{align}
Here $c^{(m)}_\pm$ and $d^{(m)}_\pm$ are integration constants from the view points of  the differential equations 
but they should satisfy the continuity of the evolution coming from the radiation dominated epoch.
And $\beta$ is defined by
\begin{align}
\beta \equiv \sqrt{-48 \eta  \xi +96 \xi  \omega +81} \,. \label{condbeta}
\end{align}

We see that the solution of temporal component of gauge field have two different type of evolution depending on whether $\beta$ is real or imaginary.
So if the term, $-48 \eta  \xi +96 \xi  \omega +81$, inside the square root is a positive real number,
the corresponding solution of the gauge field will evolve as a power law
given by growing~($\beta > 3$) or decaying~($\beta<3$) modes.
On the other hand, if the term inside the square root is negative, the gauge field will oscillate with an amplitude proportional to $a^{-3/4}$,
\begin{align}
f(a) \simeq {\left(c^{(m)}_-+c^{(m)}_+\right)\cos \left( {{\rm Im}(\beta)\over 4} \ln a \right)
\over a^{3/4} } \,,
\end{align}
where $c_1=c_2$ for real values of the $f(a)$.
Another possibility  remaining  is
when we have $\beta=3$.
Then, the corresponding solution will converge to a constant during the matter domination epoch
\begin{align}
f(a) \simeq \frac{c^{(m)}_-}{a^{3/2}}+c^{(m)}_+  \,.
\end{align}
Concerning the evolutions of temporal component of the massive photon field  and scalar field,  according to \eqref{energypressure},
the dark energy and pressure  are given by
\footnote{
We only consider decaying modes for the temporal vector component evolution.
Then their contributions to energy density and pressure can be neglected again because they remain substantially
smaller than the scalar contributions during the radiation dominated epoch, which is followed by the decaying mode contribution of Eq. (\ref{decaying}).
}
\begin{align}
\rho^{\rm (de)} \simeq
&{m^2 \xi \over 2}\left( d^{(m)}_- + {d^{(m)}_+ \over a^{3/2}}\right)^2 \,,\label{MD} \\
p^{\rm (de)} \simeq
&-{m^2 \xi \over 2}\left( d^{(m)}_- + {d^{(m)}_+ \over a^{3/2}}\right)^2 \,,
\end{align}
and we can also calculate the equation of state parameter  as
\begin{align}
w^{\rm (de)} \equiv {p^{\rm (de)}\over \rho^{\rm (de)}} \simeq -1\,.
\end{align}
Therefore, we note that
as the universe expands, we have
$\rho^{(\rm de)} \rightarrow (d^{(m)}_-)^2 m^2 \xi/2$ and
$p^{\rm (de)} \rightarrow -(d^{(m)}_-)^2 m^2 \xi/2$ with
$w^{\rm (de)} \rightarrow -1$,
so that the scalar component of the massive  photon   field furnishes the main source of the dark energy in this epoch.

\subsection{Dark Energy Dominated Epoch}\label{anaDE}
In this section we shall study the case in which the
late time Universe becomes dominated by the dark energy\begin{align}\label{veccondi}
\rho^{(r)} \ll \rho^{\rm (de)} \,, \quad
\rho^{(m)} \ll \rho^{\rm (de)} \,,
\end{align}
so that the Friedmann equations are given by
\begin{align}
{3\over \kappa} H^2 &\simeq \rho^{\rm (de)} \,, \label{darkdomenergy} \\
-{3\over \kappa} H^2 -{2\over \kappa} \dot{H} &\simeq p^{\rm (de)}  \label{darkdompressure} \,.
\end{align}
For the subsequent analysis, it will be convenient to introduce the following ansatz for the dark energy  density by 
\footnote{
It turns out that the evolution equations \eqref{darkdomenergy} and \eqref{darkdompressure} admit series solution in terms of inverse power of the scale factor $a$. Since the higher order terms decay rapidly with the expansion of the Universe,
we only consider the leading behavior which is sufficient for our purpose and is supported by numerical analysis.  
}
\begin{align}
\rho^{\rm (de)} \simeq  {\rho^{(\rm de)}_{*} \over a^n} \,,
\end{align}
where $n$ is a constant number which will be determined by the dynamical equations. The Hubble parameter $H$ is given by
\begin{align}
{3\over \kappa}H^2 \simeq  {\rho^{(\rm de)}_{*} \over a^n} \,,
\end{align}
where $\rho^{(\rm de)}_{*}/a^n_{*}$ is the dark energy density
when its dominance \eqref{veccondi} takes place $a=a_*$.
In such a case, the corresponding field equations are
given by\begin{align}
&a^2 f'' + \left(4-{n\over2}\right)af' + W^{(f)}  f \simeq 0 \,, \\
&a^2 \varphi'' + \left(4-{n\over2}\right)a\varphi'  + W^{(\varphi)} \varphi  \simeq 0 \,,
\end{align}
with $W^{(f)}$ and $W^{(\varphi)}$  defined by
\begin{align}
W^{(f)} &\equiv \frac{3 m^2 \xi}{\kappa \rho^{\rm (de)}_{*}}a^n-6 \xi  (\eta +4 \omega )
+n \left(3\eta\xi+6\omega\xi -\frac{3}{2}\right) \,, \\
W^{(\varphi)} &\equiv \frac{3 m^2 \xi}{\kappa \rho^{\rm (de)}_{*}}a^n + 3 (n-4) \chi
 \,.
\end{align}
For the non-zero positive values, $n>0$,
the values of  $W^{(f)}$ and $W^{(\varphi)}$ are dominated by $a^n$-term,
so we can use the approximation 
with $W \equiv \frac{3 m^2 \xi}{\kappa \rho^{(de)}_{*}}$:
\begin{align}
W^{(f)} \simeq  Wa^n
\,,\quad 
W^{(\varphi)} \simeq Wa^n
 \,.
\end{align}

\begin{figure}[ht]
\begin{center}
\scalebox{0.6}[0.6]{
\includegraphics{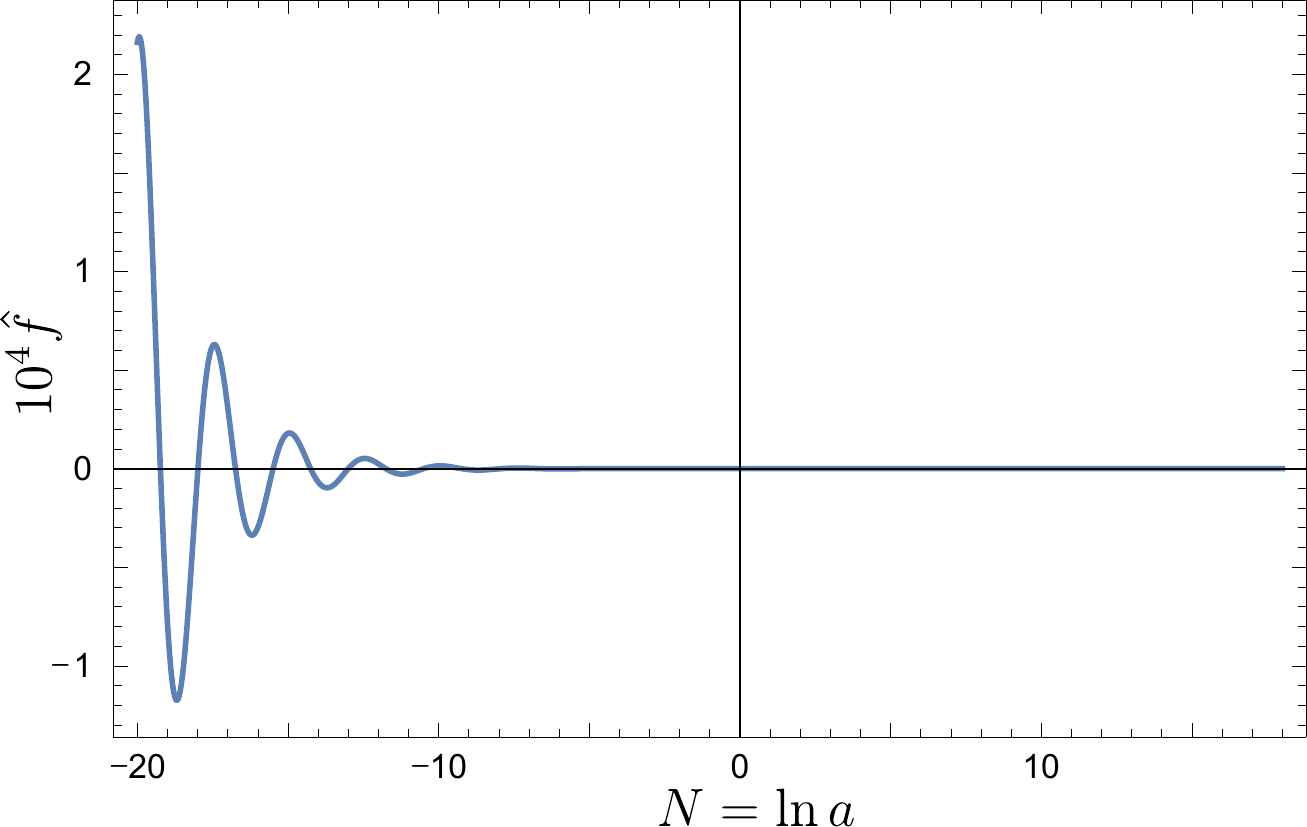}~~~
\includegraphics{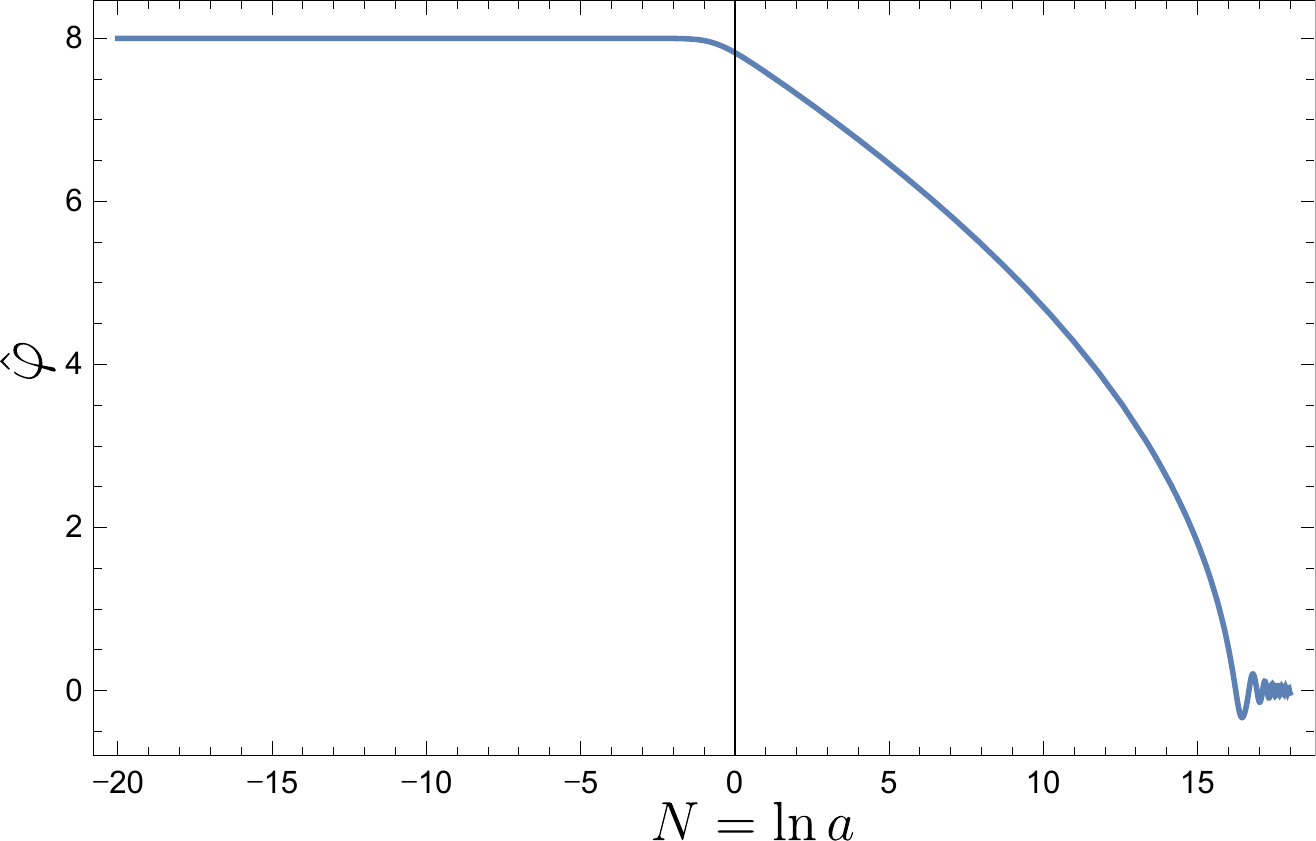}}
\end{center}
\caption{\small
Evolution of $\hat{f}$ (left) and $\hat{\varphi}$ (right) as a function of logarithmic scale factor $N=\ln a(t)$. In both panels, we have used $\hat{\eta} = 0.9$, $\hat{\omega}=-0.35$, $\chi=10^{-7}$, and $\hat{m} = 10^{-3}$. We have also set the initial values $\hat{\varphi}_i=8$ and $\hat{f}_i \simeq a_i^{(1+\sqrt{25-24 \hat{\eta}})/2} \simeq 2\times 10^{-4}$ at the initial epoch $\ln a_i=-20$.
}\label{fig:fphisol}
\end{figure}

\begin{figure}[ht]
\begin{center}
\scalebox{0.6}[0.6]{
\includegraphics{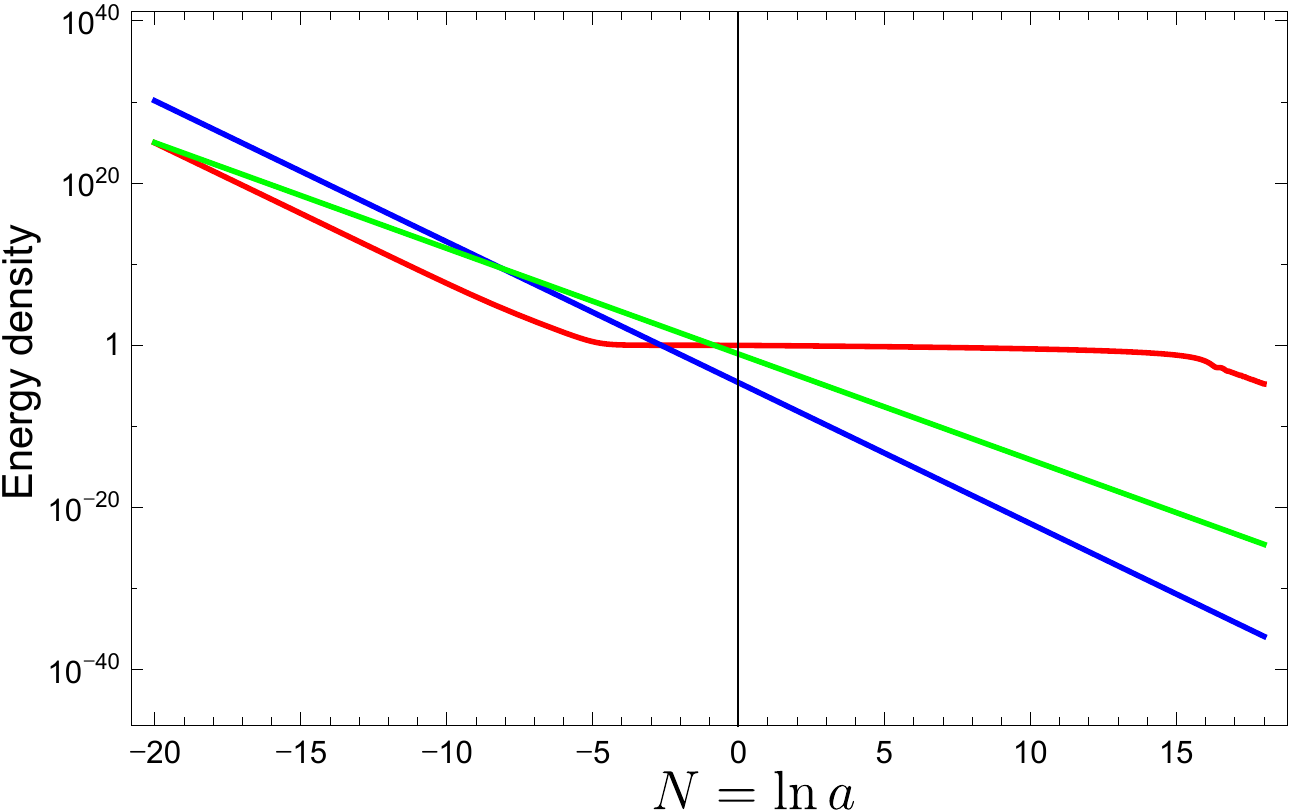}~~~
\includegraphics{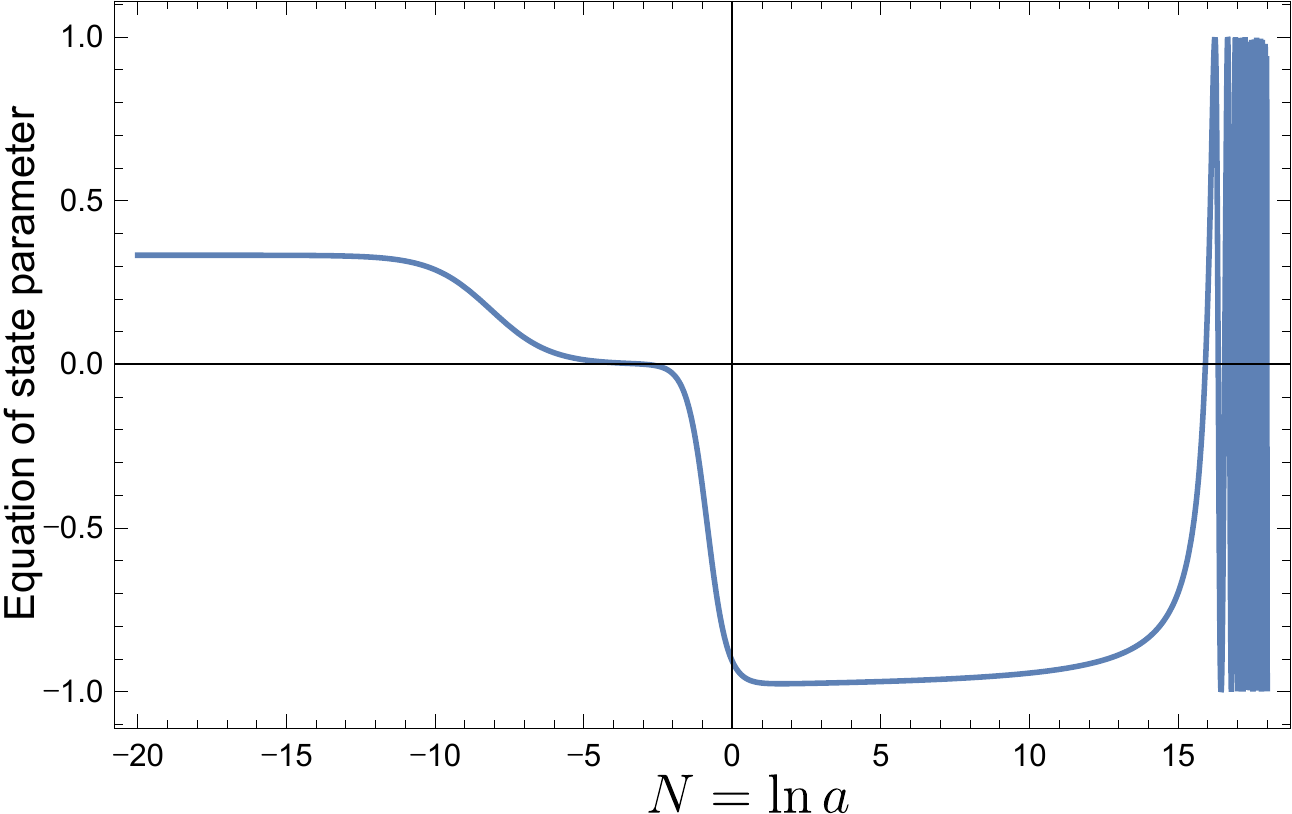}}
\end{center}
\caption{\small
(Left) Evolution of energy density of the vector field (red), radiation (blue), and matter (green curve).
(Right) Evolution of equation of state parameter. The same model parameters have been used as in Fig.~\ref{fig:fphisol}.
}\label{fig:energyEOS}
\end{figure}

The corresponding solutions are given by
\begin{align}
&f \simeq  \frac{1}{a^{3/2}}\Biggr[
c^{(v)}_- \cos \left(\frac{2 \sqrt{W}}{n} a^{n/2}-\frac{3 \pi }{2 n}\right)
+c^{(v)}_+ \sin \left(\frac{2 \sqrt{W}}{n} a^{n/2}+\frac{3 \pi }{2 n}\right)
\Biggr] \,, \label{vecsolf} \\
& \varphi \simeq \frac{1}{a^{3/2}}\Biggr[
d^{(v)}_- \cos \left(\frac{2 \sqrt{W}}{n} a^{n/2}-\frac{3 \pi }{2 n}\right)
+d^{(v)}_+ \sin \left(\frac{2 \sqrt{W}}{n} a^{n/2}+\frac{3 \pi }{2 n}\right)
\Biggr] \,, \label{vecsolphi}
\end{align}
where $c^{(v)}_{\pm}$ and $d^{(v)}_{\pm}$ are integration constant.
Here, we note that when we insert the above solutions into \eqref{energypressure},
Eq. \eqref{veccondi} forces  $n=3$,
and thus the corresponding energy density and pressure of the massive vector field are given by
\begin{align}
&\rho^{\rm (de)} \simeq \frac{m^2\xi/2 }{a^3}
\left[
(d^{(v)}_-)^2+(d^{(v)}_+)^2 -{ (c^{(v)}_-)^2+(c^{(v)}_+)^2 \over \xi}
\right]
\,, \label{vecenergy}\\
& p^{\rm (de)} \simeq \frac{m^2 \xi /2 }{a^3}
\left[
p_{+} \sin \left(\frac{4}{3} a^{3/2} \sqrt{W}\right)
+p_{-}\cos \left(\frac{4}{3} a^{3/2} \sqrt{W}\right)
\right]
\,, \label{vecpressure}
\end{align}
where $p_+$ and $p_-$ are the amplitudes  given by
\begin{align}
&p+ \equiv  2 c^{(v)}_- c^{(v)}_+ \left(4 \eta  +8  \omega -{1\over \xi} \right)
-2 d^{(v)}_-  d^{(v)}_+ \,, \\
&p_- \equiv \Big((c^{(v)}_-)^2-(c^{(v)}_+)^2\Big) \left(-4 \eta -8  \omega +{1\over \xi} \right)
+(d^{(v)}_-)^2-(d^{(v)}_+)^2 \,.
\end{align}
Note that 
if \eqref{vecenergy} is exactly correct then the corresponding pressure
should be  zero. But it shows only the leading behavior in the expansion.
If we calculate the next order, for example, $\rho^{\rm (de)}$ will be augmented by an oscillating term whose magnitude decays as power of $\sim 1/a^{9/2}$ 
as was mentioned before.
Then,  the equation of state parameter for the dark energy is given by
\begin{align}
\omega^{\rm (de)} = {p^{\rm (de)}\over \rho^{de}} \simeq
{p_{+} \sin \left(\frac{4\sqrt{W}}{3} a^{3/2} \right)
+p_{-}\cos \left(\frac{4\sqrt{W}}{3} a^{3/2} \right)
\over
(d^{(v)}_-)^2+(d^{(v)}_+)^2 -{(c^{(v)}_-)^2+(c^{(v)}_+)^2 \over \xi}}
  \,.
\end{align}
We  note that the equation of state parameter has oscillation terms
which gives zero average value.
Thus, the corresponding energy density should be proportional to $1/a^3$,
which is consistent with energy density equation~\eqref{vecenergy}.
We plot behaviors of $f$ and $\varphi$ based on numerical solution
to confirm  our analytically approximated solution, in Fig.~\ref{fig:fphisol}.
The Fig.~\ref{fig:energyEOS}
shows that the dark energy density decreases as $a^{-4}$ during the early radiation-dominated epoch,
remains almost constant during the matter-dominated epoch
and then it decreases again as $a^{-3}$ in the dark energy dominated era.

%

\newpage
\section{Observational Constraints}\label{secob}
In this section we will confront our model
with the latest cosmological data and study whether
it can be distinguished from the $\Lambda$-CDM model.
For this purpose,
we use the recent observational data such as type Ia supernovae (SN),
baryon acoustic oscillation (BAO) based on large-scale structure of galaxies,
cosmic microwave background radiation (CMB), and Hubble parameters [$H(z)$].
For numerical analysis, 
it is convenient to rewrite equations~\eqref{eqgauge}-\eqref{eqH}
in terms of
$N\equiv \ln a$ as follows:
\begin{align}\label{eqH2com}
\hat{H}^2
&=\frac{1}{6} \hat{m}^2 \hat{\varphi} ^2
-\frac{1}{6}\hat{m}^2 \hat{f}^2
\nonumber \\
& + {\hat{H}^2\over 3} \Biggr[
\hat{f}^2 \left(6 \hat{\eta} -\frac{9}{2}+6 \hat{\omega} \right)
-3 \chi\hat{\varphi} ^2  +{1\over 2} \left(\hat{\varphi}'^2-\hat{f}'^2\right) + \hat{f}\hat{f}' \left(6 \hat{\eta} -3+12 \hat{\omega} \right)
-6  \chi   \hat{\varphi}\hat{\varphi}'
\Biggr] \nonumber \\
&+\Omega_r h^2 e^{-4N}
+\Omega_m h^2 e^{-3N}
\,,
\end{align}
where a prime indicates a derivative with respect to $N$, and
\begin{align}
&\hat{H}^2 \hat{f}''
+ \left(\hat{H} \hat{H}'+3 \hat{H}^2\right)\hat{f}'
+\left[ \hat{m}^2
-3\hat{H}\hat{H}' \left(-1+2\hat{\eta}+4\hat{\omega}\right)
-6\hat{H}^2 \left( \hat{\eta}+4\hat{\omega}\right)
\right] \hat{f}
=0 \nonumber \,, \\
&\hat{H}^2 \hat{\varphi} ''
+\left(\hat{H} \hat{H}'+3\hat{H}^2\right)\hat{\varphi} '
+  \left(\hat{m}^2 -6 \chi \hat{H} \hat{H}'-12 \chi  \hat{H}^2\right) \hat{\varphi}
=0
\,,
\end{align}
where we have eliminated the second order derivative in \eqref{eqH2com} by using the field equations,
and have introduced dimensionless quantities,
\begin{align}
\hat{H}^2 &\equiv {H^2 h^2\over H_0^2} \,, \quad
\Omega_r \equiv {\kappa \rho_{r,0}\over 3H_0^2}\,, \quad
\Omega_m \equiv {\kappa \rho_{m,0}\over 3H_0^2}\,, \quad
\hat{m}^2 \equiv {m^2 \xi h^2 \over H_0^2} \,,
\nonumber \\
\hat{f} &\equiv {\kappa f \over \xi}  \,, \quad
~~~~~\hat{\varphi} \equiv \kappa \varphi \,, \quad
~~~~~\hat{\eta} = \eta \xi \,, \quad
~~~~~~~\hat{\omega} = \omega \xi\,.
\end{align}
Here, $H_0$ is the present value of the Hubble parameter, 
usually expressed as $H_0=100 \,h\, {\rm km}\,\,{\rm s}^{-1}{\rm Mpc}^{-1}$,
$\Omega_r$ and $\Omega_m$ are the current density parameters of radiation and matter, respectively. 
The radiation density includes the contribution of relativistic neutrinos as well as that of photons, 
with the collective density parameter
\begin{align}
\Omega_r h^2 = \Omega_\gamma h^2 \left(1+0.2271N_{\rm eff}\right)
\,,
\end{align}
where $N_{\rm eff}=3.04$ is the effective number of neutrino species, and
$\Omega_\gamma$ is the photon density parameter with values of $\Omega_r=2.47037\times 10^{-5} h^{-2}$ for the present CMB temperature $T_0=2.725~{\rm K}$~(WMAP9) and $\Omega_r=2.47218\times 10^{-5} h^{-2}$ for $T_0=2.7255~{\rm K}$~(PLANCK).
Notice that, in this analysis,
we shall choose the decaying mode for the vector field $\hat{f}$ during the matter era, 
which satisfies a condition
$\sqrt{-48 \hat{\eta} +96\hat{\omega} +81} < 3$ in \eqref{condbeta}.
In such a case, the contribution of the temporal component is negligible
relative to the scalar field, and the decaying-mode condition gives almost the same probability in the parameter constraints for $\hat\eta$ and $\hat\omega$. Thus, we choose $\hat\eta=0.9$ and $\hat\omega=-0.35$ as fixed values during our analysis.\footnote{
There are strong constraints from local gravity experiments which, among others, imply
a small value for the $\omega$ parameter.
In our case $f^2 \omega \ll 1$ and the PPN $\gamma$ parameter~\cite{Will:1993ns}
 is very close to unity which does not cause a enough change of the gravitational constant
 to be incompatible with the observation, that is, $\vert \gamma-1\vert<2\times 10^{-5}$~ \cite{Bertotti:2003rm}.
}
Therefore the background dynamics is completely determined by a set of parameters
$(\hat{m}, \chi, \hat{\varphi}_i, \Omega_m)$.
However, to confront our model with the real observational data, we need an additional parameter of baryon density ($\Omega_b$), and finally our model has five free parameters 
$\mbox{\boldmath $\theta$}=(\log_{10} \hat{m}, -\log_{10}(-\chi),\log_{10} {\hat{\varphi}}_i, \Omega_b h^2, \Omega_m h^2)$.
It should be emphasized that the Hubble constant ($H_0$) is no longer a free parameter
because it is derived from the integration of field equations for a given set of parameters chosen.
The free parameters are taken in the following priors:
$\log_{10} \hat{m} = [-3, 3]$, $-\log_{10} (-\chi) = [1,7]$,
$\log_{10} \hat{\varphi}_i = [-3, 3]$, $\Omega_b h^2 = [0.015, 0.030]$
and $\Omega_m h^2 = [0.11, 0.15]$.
In addition, as mentioned above, for the analysis,
we fixed the parameters as $\hat{\eta}=0.9$ and $\hat{\omega}=-0.35$.
To obtain the likelihood distributions for model parameters, we use the
Markov chian Monte Carlo (MCMC) method based on Metropolis-Hastings
algorithm to randomly explore the parameter space that is favored by
observational data \cite{MCMC}.
The method needs to make decisions for accepting or rejecting
a randomly chosen chain element via the probability function
$P(\mbox{\boldmath $\theta$}|\mathbf{D}) \propto \exp(-\chi^2/2)$,
where $\mathbf{D}$ denotes the data,
and $\chi^2= \chi_{H(z)}^2+\chi_\textrm{SN}^2 + \chi_\textrm{BAO}^2 + \chi_\textrm{CMB}^2
$ is the sum of individual chi-squares
for $H(z)$, SN, BAO, and CMB data (defined below).
During the MCMC analysis, we use a simple diagnostic to test the convergence
of MCMC chain: the means estimated from the first (after buring process)
and the last 10\% of the chain are approximately equal to each other
if the chain has converged (see Appendix B of Ref.\ \cite{Abrahamse-etal-2008}).

\subsection{Hubble Parameters}
In our analysis, we use 29 observational data points of Hubble parameters over a redshift range of
$0.07 \leq z \leq 2.34$, which include 23 data points obtained from the differential age approach~\cite{Hubble:DA} and 6 derived from the BAO measurements~\cite{Hubble:BAO}. The chi-square is defined as
\begin{align}
\chi^2_{H(z)} = \sum_{i=1}^{29}
\frac{ \left[H_{\textrm{th}}(z_i) - H_{\textrm{obs}}(z_i)  \right]^2}{\sigma^2_H(z_i)}
\,,   
\end{align}
where $H_{\textrm{th}}(z_i)$ and $H_{\textrm{obs}}(z_i)$ are theory-predicted and observed
values of the Hubble parameter at redshift $z_i$, respectively, and $\sigma_H$ denotes the measurement error
of the observed data point.

\subsection{Type Ia Supernovae}
The type Ia supernovae provide tight constraints on the energy content of the late-time Universe.
We use the Union 2.1 compilation~\cite{Suzuki:2011hu} that includes 580 SNe over a redshift range of $0.015 \leq z \leq 1.414$.
In our analysis, 
we apply the chi-square that has been marginalized over the zero-point uncertainty
due to absolute magnitude and Hubble constant \cite{SNIa-MRG}:
\begin{align}
    \chi_{\textrm{SN}}^2 = c_1 - c_2^2 / c_3,
\end{align}
where
\begin{align}
     c_1 = \sum_{i=1}^{580}
       \left[ \frac{\mu_{\rm th}(z_i)-\mu_\textrm{obs}(z_i)}{\sigma_i}\right]^2,
      \quad
     c_2= \sum_{i=1}^{580}
        \frac{\mu(z_i)_{\rm th}-\mu_\textrm{obs}(z_i)}{\sigma_i^2},
       \quad
     c_3= \sum_{i=1}^{580} \frac{1}{\sigma_i^2},
\end{align}
where $\mu_\textrm{obs}(z_i)$ and $\sigma_i$ denote the observed distance modulus and its measurement error of SN at redshift $z_i$. The theoretical distance modulus $\mu_{\rm th}$ is defined as
\begin{align}
\mu_{\rm th}(z)=5\log [(1+z) r(z)]\,,
\end{align}
where $r(z)$ is the comoving distance at redshift $z$,
\begin{align}
   r(z)=\frac{c}{H_0 \sqrt{\Omega_k}}
       \sin \left[\sqrt{\Omega_k}\int_0^z \frac{H_0}{H(z')} dz' \right],
\end{align}
with $c$ the speed of light and $\Omega_k$ the current density parameter of spatial curvature
($\Omega_k=0$ in our analysis).

\subsection{Baryon Acoustic Oscillations}
We use an effective distance measure which is related to the BAO scale
\cite{Eisenstein-etal-2005},
\begin{align}
   D_V(z) \equiv \left[r^2 (z) \frac{cz}{H(z)} \right]^\frac{1}{3}\,,
\end{align}
and a fitting formula for the redshift of drag epoch ($z_d$)
\cite{Eisenstein-Hu-1998}:
\begin{align}
   z_d = \frac{1291(\Omega_m h^2)^{0.251}}{1+0.659(\Omega_m h^2)^{0.828}}
       \left[1 + b_1 (\Omega_b h^2)^{b_2} \right],
\end{align}
where
\begin{align}
   b_1 =0.313 (\Omega_m h^2)^{-0.419}
       \left[1+0.607(\Omega_m h^2)^{0.674} \right], \quad
   b_2 =0.238 (\Omega_m h^2)^{0.223}.
\end{align}
As the BAO parameter, we use six numbers of $r_s(z_d)/D_V(z)$ extracted
from the Six-Degree-Field Galaxy Survey \cite{6dFGS},
the Sloan Digital Sky Survey Data Release 7 and 9 \cite{SDSS},
and the WiggleZ Dark Energy Survey \cite{WiggleZ},
where $r_s(z)$ is the comoving sound horizon size.
These BAO data points were used in the WMAP 9-year analysis
\cite{WMAP9}.
Since the sound speed of baryon fluid coupled with photons ($\gamma$)
is given as
\begin{align}
   c_s^2 = \frac{\dot{p}}{\dot\rho}
         = \frac{\frac{1}{3}\dot\rho_\gamma}{\dot\rho_\gamma + \dot\rho_b}
         = \frac{1}{3\left[1+(3\Omega_b/4\Omega_\gamma)a\right]},
\end{align}
the comoving sound horizon size before the last scattering becomes
\begin{align}
   r_s(z) = \int_0^t c_s dt'/a
          = \frac{1}{\sqrt{3}} \int_0^{1/(1+z)}
               \frac{da}{a^2 H(a)[1+(3\Omega_b/4\Omega_\gamma)a]^\frac{1}{2}} \,.
\end{align}
The BAO measurements provide the following distance ratios~\cite{WMAP9}
\begin{align}
&\left<r_s(z_d)/D_V(0.1)\right> =0.336\,,\quad~~\,
\left<D_V(0.35)/r_s(z_d)\right> =8.88\,,\\
&\left<D_V(0.57)/r_s(z_d)\right> =13.67\,, \quad~
\left<r_s(z_d)/D_V(0.44)\right> =0.0916\,,\\
&\left<r_s(z_d)/D_V(0.60)\right> =0.0726\,, \quad
\left<r_s(z_d)/D_V(0.73)\right> =0.0592 \,.
\end{align}
The inverse of the covariance matrix between measurement errors is
\begin{align}
    \mathbf{C}_{\textrm{BAO}}^{-1} =
          \left( \begin{array}{cccccc}
          4444.4 & 0 & 0 & 0 & 0 & 0  \\
             0   & 34.602 & 0 & 0 & 0 & 0  \\
             0   & 0 & 20.661157 & 0 & 0 & 0  \\
             0   & 0 & 0 & 24532.1 & -25137.7 & 12099.1  \\
             0   & 0 & 0 & -25137.7 & 134598.4 & -64783.9  \\
             0   & 0 & 0 & 12099.1 & -64783.9 & 128837.6 \end{array} \right).
\end{align}
The chi-square is given as
\begin{align}
\chi_{\textrm{BAO}}^2
=\mathbf{X}^T
\mathbf{C}_{\textrm{BAO}}^{-1} \mathbf{X}
\,,
\end{align}
where
\begin{align}
\mathbf{X}=
\left( \begin{array}{c}
             r_s(z_d)/D_V(0.1) - 0.336\\
             D_V(0.35)/r_s(z_d)  - 8.88\\
             D_V(0.57)/r_s(z_d) - 13.67\\
             r_s(z_d)/D_V(0.44) - 0.0916\\
             r_s(z_d)/D_V(0.60) - 0.0726\\
             r_s(z_d)/D_V(0.73) - 0.0592 \end{array} \right)
\,.
\end{align}

\subsection{Cosmic Microwave Background Radiation}
As the CMB data,
we use the CMB distance priors based on WMAP 9-year data~\cite{WMAP9}
and Planck data~\cite{Ade:2013zuv} for testing our model.
The first distance measure is the acoustic scale $l_A$ defined as
\begin{align}
   l_A = \pi \frac{r(z_*)}{r_s (z_*)}.
\end{align}
The decoupling epoch $z_*$ can be calculated from the fitting function
\cite{Hu-Sugiyama-1996}:
\begin{align}
   z_*=1048 [1+0.00124(\Omega_b h^2)^{-0.738}]
           [1+g_1(\Omega_m h^2)^{g_2}]\,,
\end{align}
where
\begin{align}
   g_1 = \frac{0.0783(\Omega_b h^2)^{-0.238}}{1+39.5(\Omega_b h^2)^{0.763}},
   \quad
   g_2 = \frac{0.560}{1+21.1(\Omega_b h^2)^{1.81}}.
\end{align}
The second distance measure is the shift parameter $R$ which is given by
\begin{align}
   R(z_*) = \frac{\sqrt{\Omega_m H_0^2}}{c} r(z_*).
\end{align}
Recently, Shafer \& Huterer \cite{Shafer:2013pxa} derived distance priors from the WMAP and Planck data and provided mean values and covariance matrix of the parameter combination $(l_a,R,z_{*})$ as an efficient summary of CMB information on dark energy. Hereafter, we use these data sets to constrain our model parameters.

\subsubsection{WMAP 9-year data}
According to WMAP 9-year observations (WMAP9) \cite{WMAP9},
the mean values for the three parameters
$(l_A, R, z_*)$ are given as \cite{Shafer:2013pxa}
\begin{align}
\left<l_A (z_*)\right> =301.98\,,\quad
\left<R(z_*)\right> =1.7302\,, \quad
\left<z_* \right> =1089.09\,,  
\end{align}
with their inverse covariance matrix
\begin{align}
     \mathbf{C}_{\textrm{WMAP9}}^{-1} =    
\left(
\begin{array}{ccc}
 3.13365 & 15.1332 & -1.43915 \\
 15.1332 & 13343.7 & -223.16 \\
 -1.43915 & -223.16 & 5.44598 \\
\end{array}
\right).
\end{align}
The chi-square is given as
\begin{align}
\chi_{\textrm{WMAP9}}^2
=\mathbf{X}^T
\mathbf{C}_{\textrm{WMAP9}}^{-1} \mathbf{X}
\,,
\end{align}
where
\begin{align}
    \mathbf{X}=
       \left( \begin{array}{c}
             l_A (z_*) - 301.98 \\
             R(z_*) - 1.7302 \\
             z_*    - 1089.09 \end{array} \right)
             \,.
\end{align}

\subsubsection{Planck data}

According to Planck observations (PLANCK) \cite{Ade:2013zuv},
the mean values for the three parameters
$(l_A, R, z_*)$ are given as \cite{Shafer:2013pxa}
\begin{align}
\left<l_A (z_*)\right> =301.65\,,\quad
\left<R(z_*)\right> =1.7499\,, \quad
\left< z_* \right> =1090.41\,.
\end{align}
Their inverse covariance matrix is
\begin{align}
 \mathbf{C}_{\textrm{Planck}}^{-1} =    
\left(
\begin{array}{ccc}
 42.7223 & -419.678 & -0.765895 \\
 -419.678 & 57394.2 & -762.352 \\
 -0.765895 & -762.352 & 14.6999 \\
\end{array}
\right).
\end{align}
The chi-square is given as
\begin{align}
\chi_{\textrm{Planck}}^2
=\mathbf{X}^T
\mathbf{C}_{\textrm{Planck}}^{-1} \mathbf{X}
\,,
\end{align}
where
\begin{align}
    \mathbf{X}=
       \left( \begin{array}{c}
             l_A (z_*) - 301.65 \\
             R(z_*) - 1.7499 \\
             z_*    - 1090.41 \end{array} \right)
             \,.
\end{align}


\subsection{Results}

\begin{figure}
\begin{minipage}[t]{0.32\textwidth}
\includegraphics[width=\linewidth]{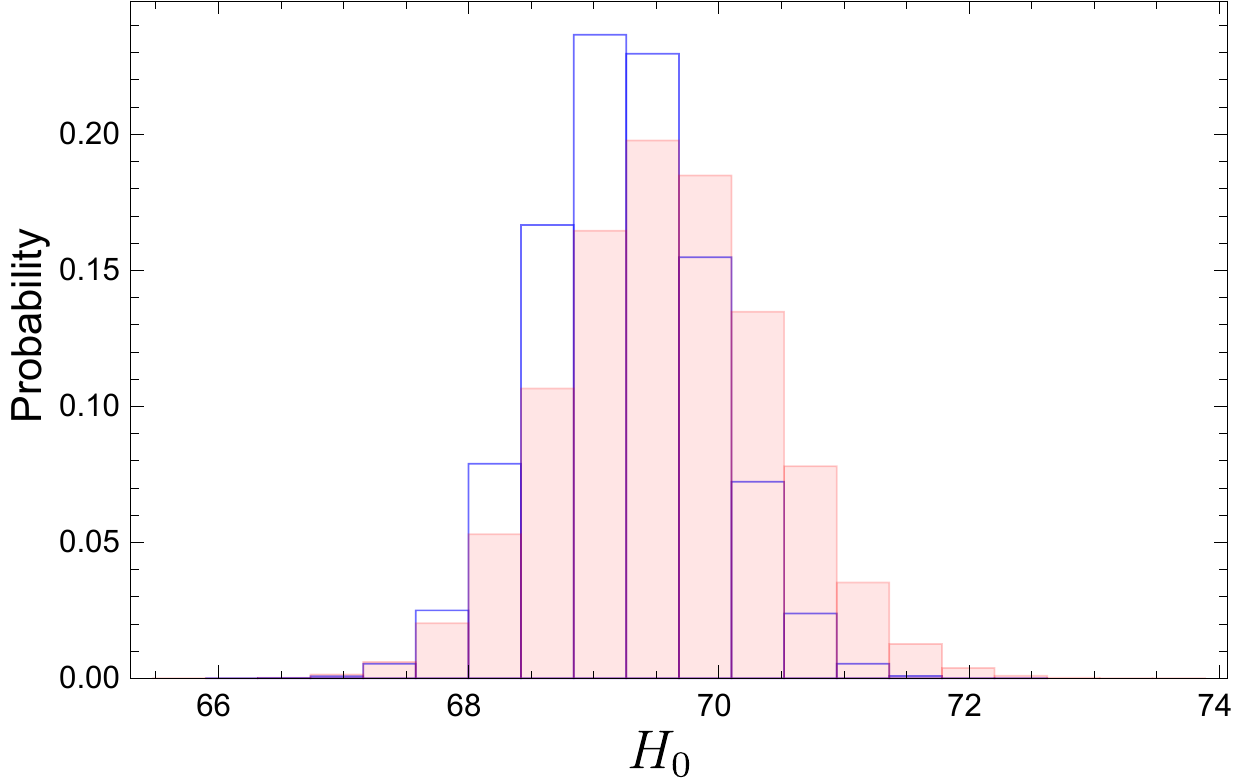}
\end{minipage}
\hspace{\fill}
\begin{minipage}[t]{0.32\textwidth}
\includegraphics[width=\linewidth]{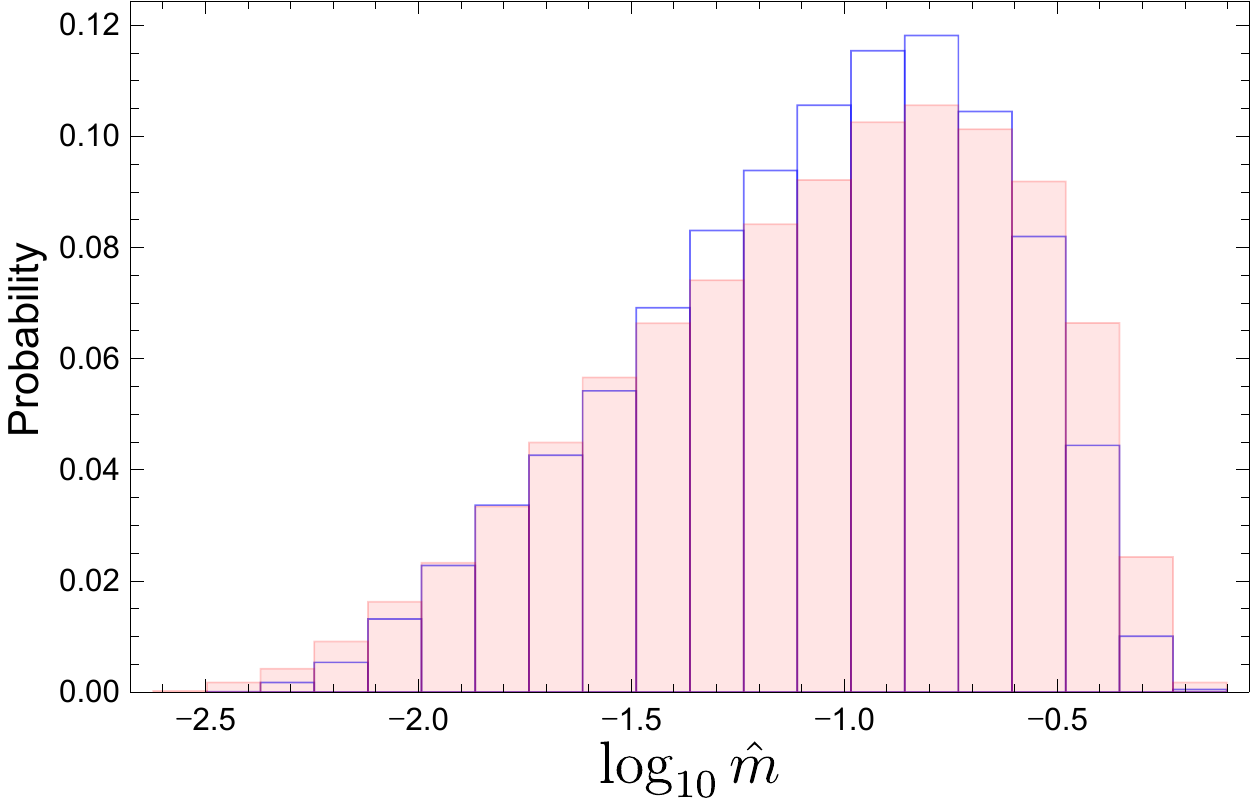}
\end{minipage}
\hspace{\fill}
\begin{minipage}[t]{0.32\textwidth}
\includegraphics[width=\linewidth]{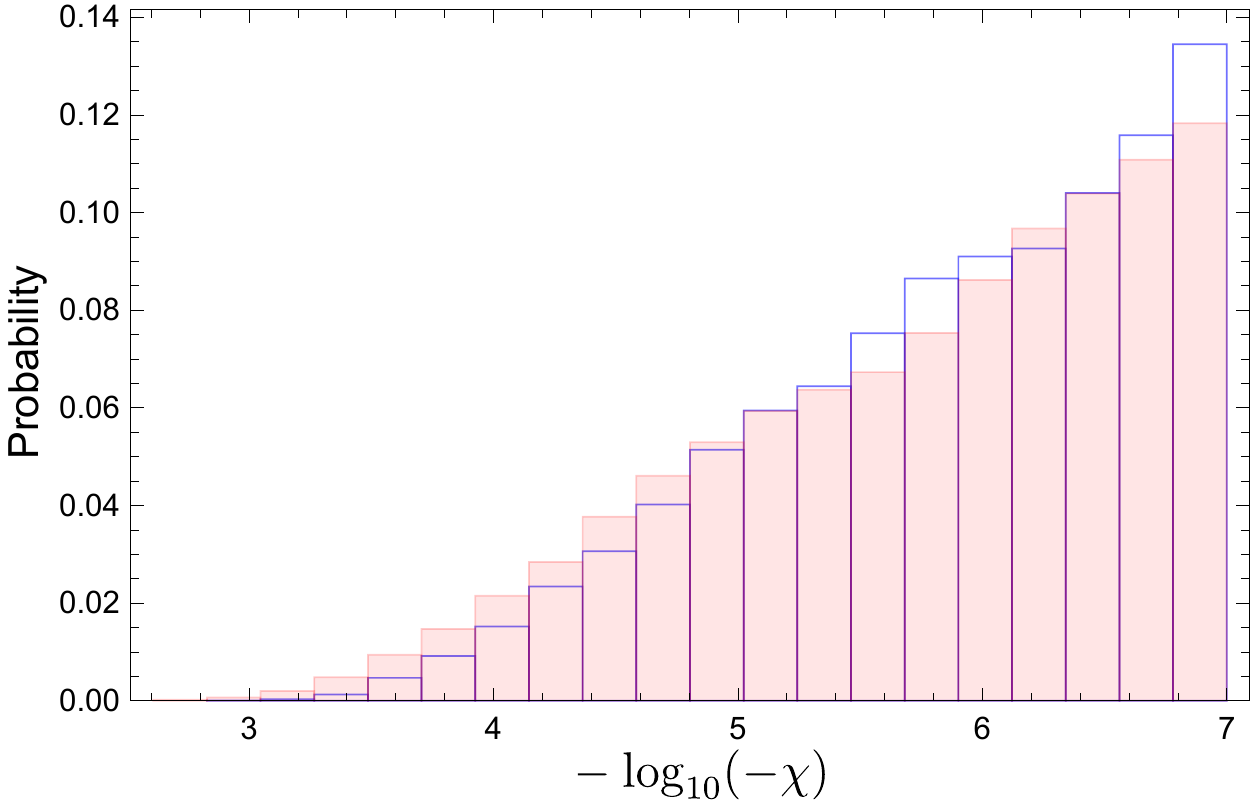}
\end{minipage}

\vspace*{0.5cm}
\begin{minipage}[t]{0.32\textwidth}
\includegraphics[width=\linewidth]{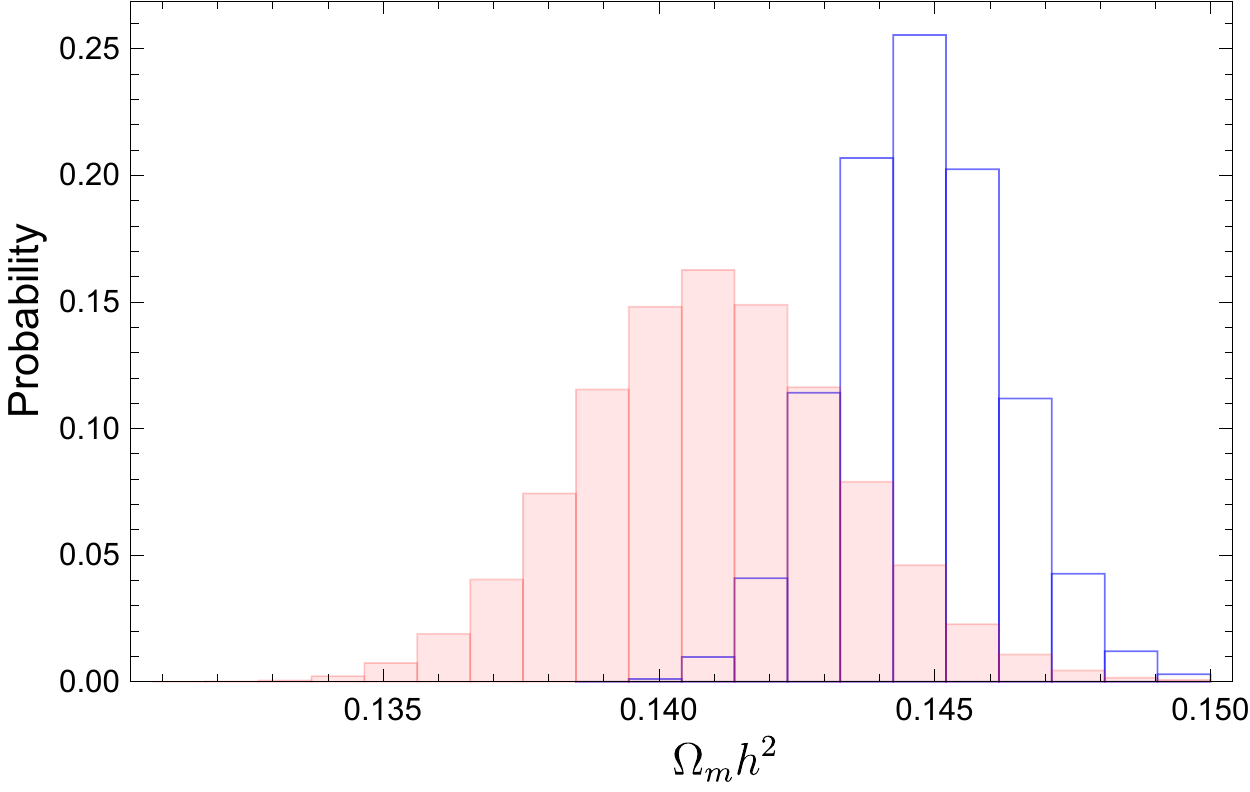}
\end{minipage}
\hspace{\fill}
\begin{minipage}[t]{0.32\textwidth}
\includegraphics[width=\linewidth]{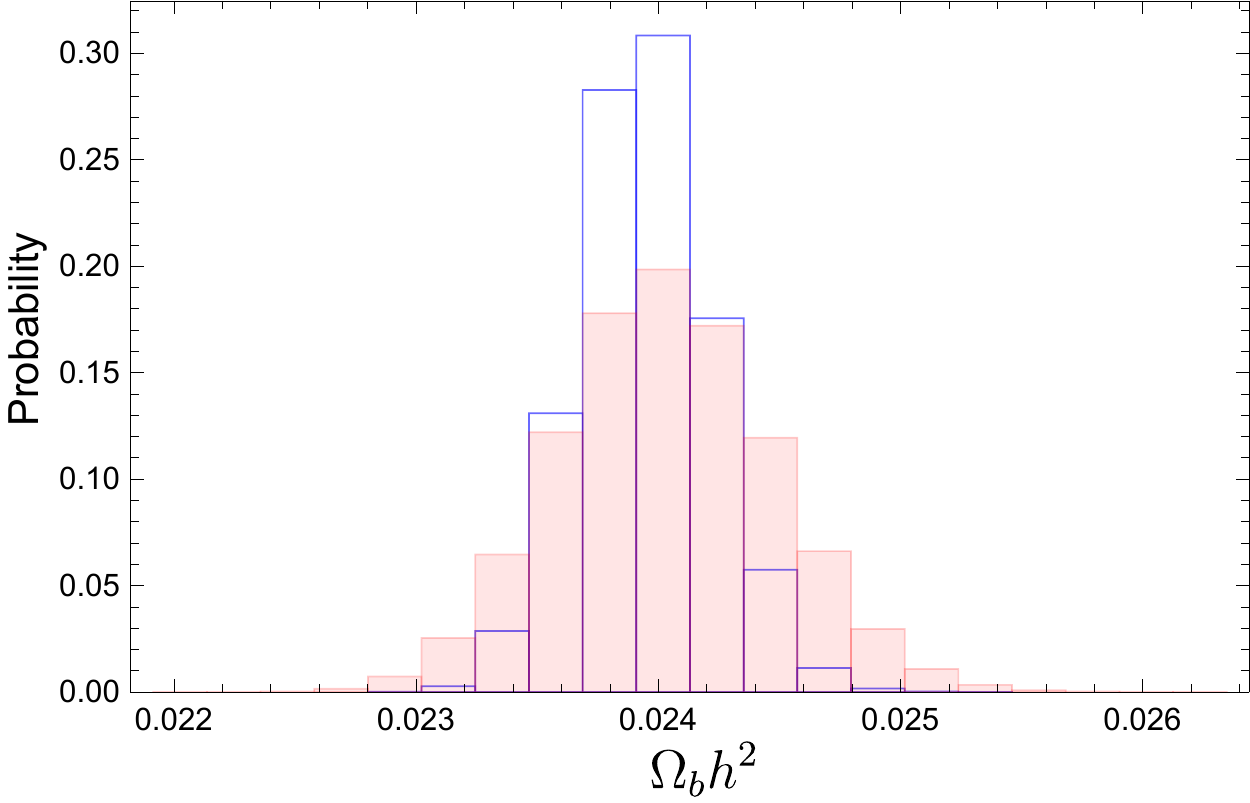}
\end{minipage}
\hspace{\fill}
\begin{minipage}[t]{0.32\textwidth}
\includegraphics[width=\linewidth]{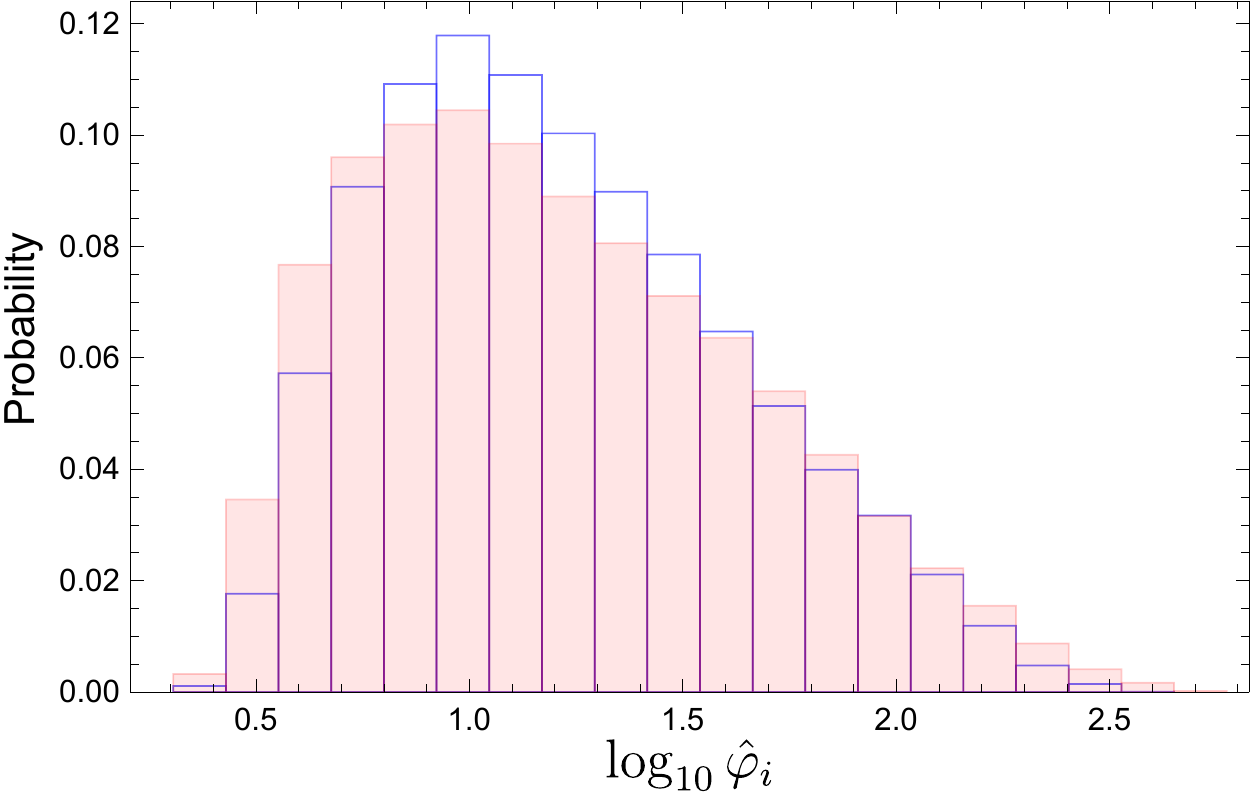}
\end{minipage}
\caption{\small
Marginalized one-dimensional probability distributions of Hubble constant ($H_0$) and five model parameters
($\log_{10}\hat{m}$, $-\log_{10}(-\chi)$, $\Omega_m h^2$, $\Omega_b h^2$, $\log_{10}\hat{\varphi}_i$), favored by the current
observations; $H(z)$+SN+BAO+PLANCK (blue) and $H(z)$+SN+BAO+WMAP9 (red histograms), respectively.
}
\label{fig:like}
\end{figure}

\begin{figure}[ht]
\begin{center}
\scalebox{0.6}[0.6]{
\includegraphics{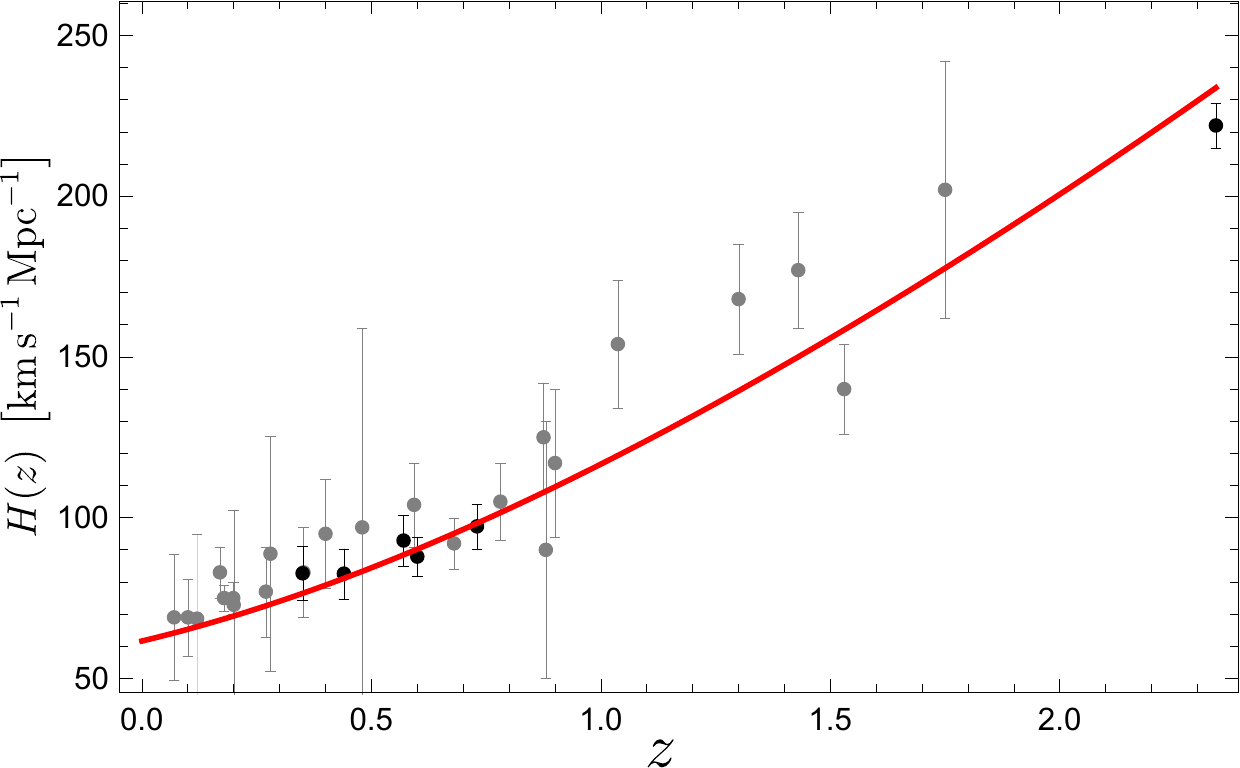}~~\includegraphics{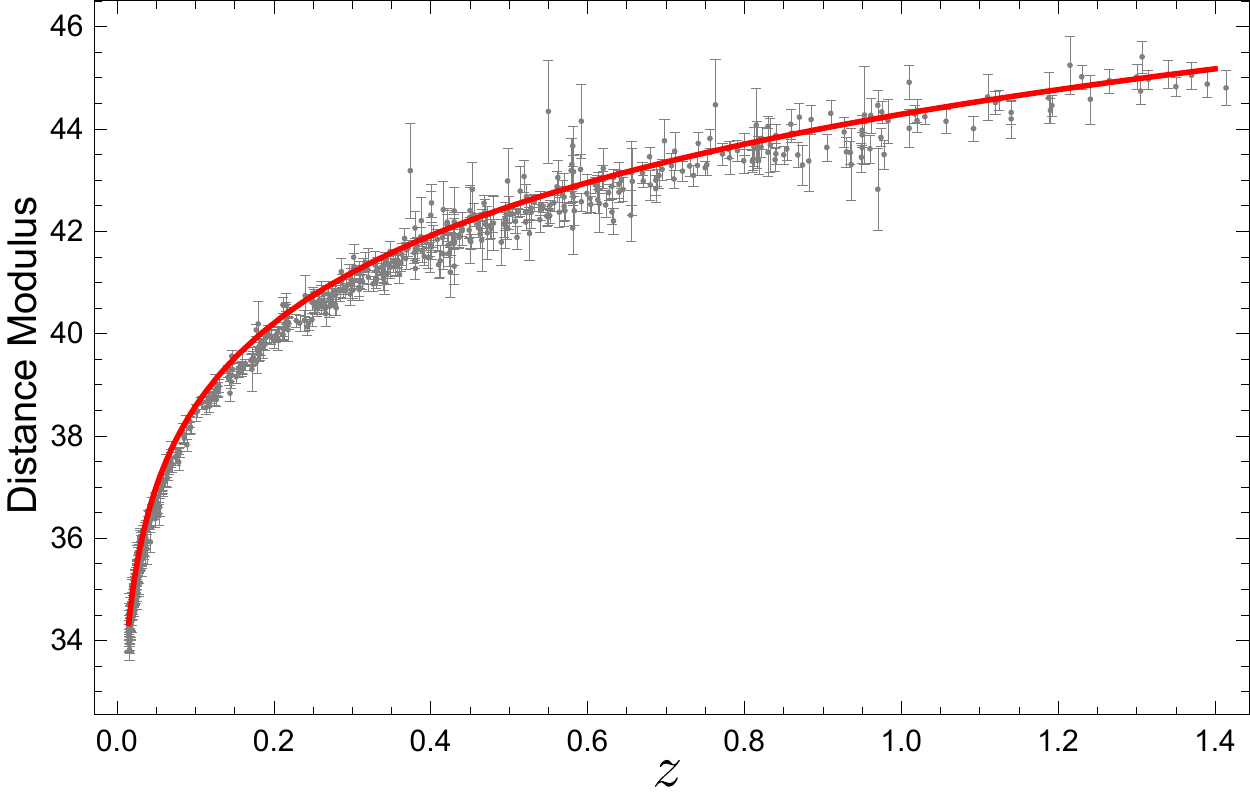}}
\end{center}
\caption{\small
(Left) Observed Hubble parameters versus redshift (grey and black dots with error bars; see text). (Right)
The Hubble diagram for Union 2.1 compilation of SNe Ia. In both figures, the red curve represents the best-fit prediction
of our model constrained with $H(z)$, SN, BAO, and CMB data sets.
}\label{fig:bestH}
\end{figure}

\begin{figure}[ht]
\begin{center}
\scalebox{0.55}[0.55]{
\includegraphics{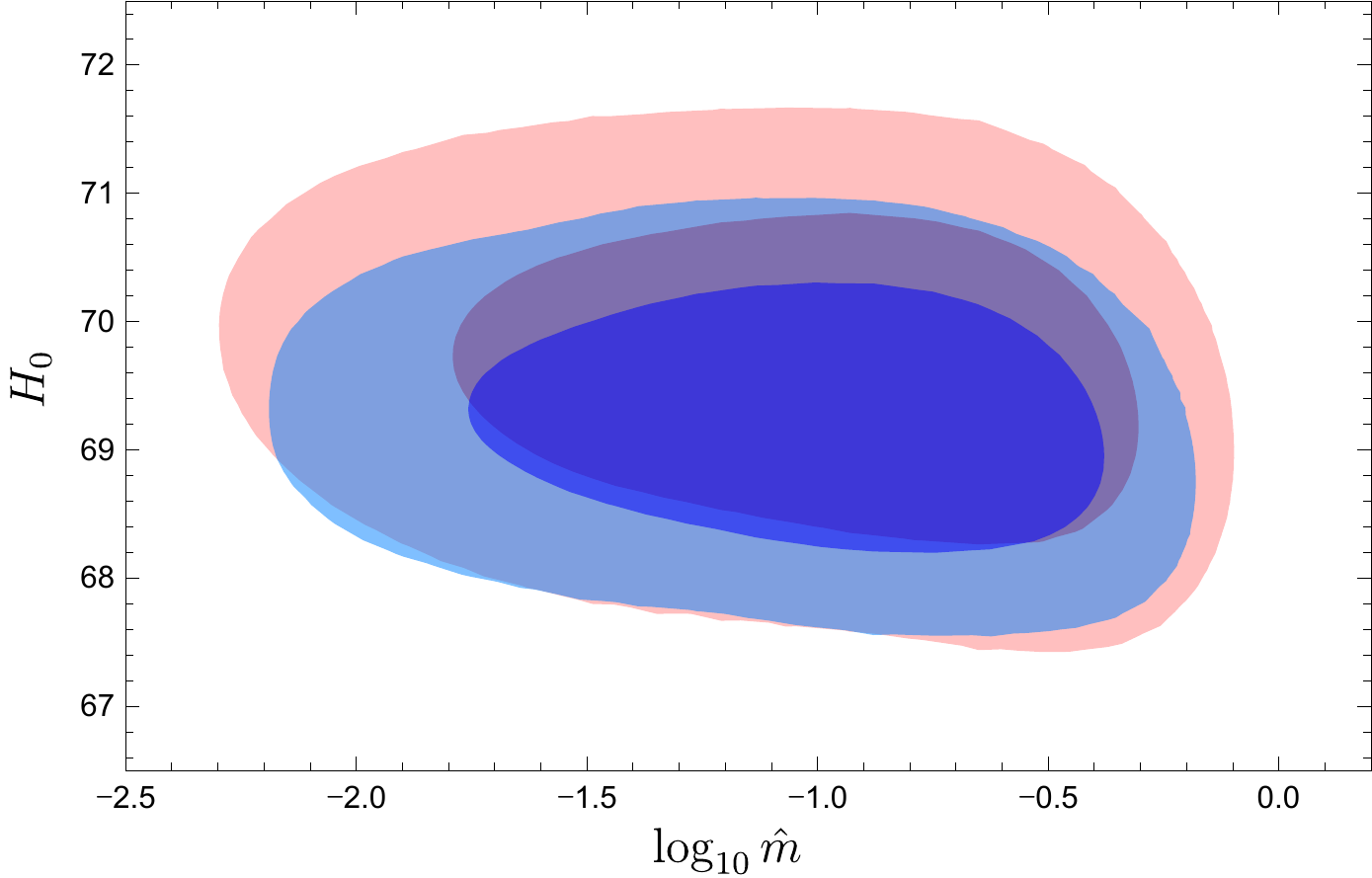}~~~\includegraphics{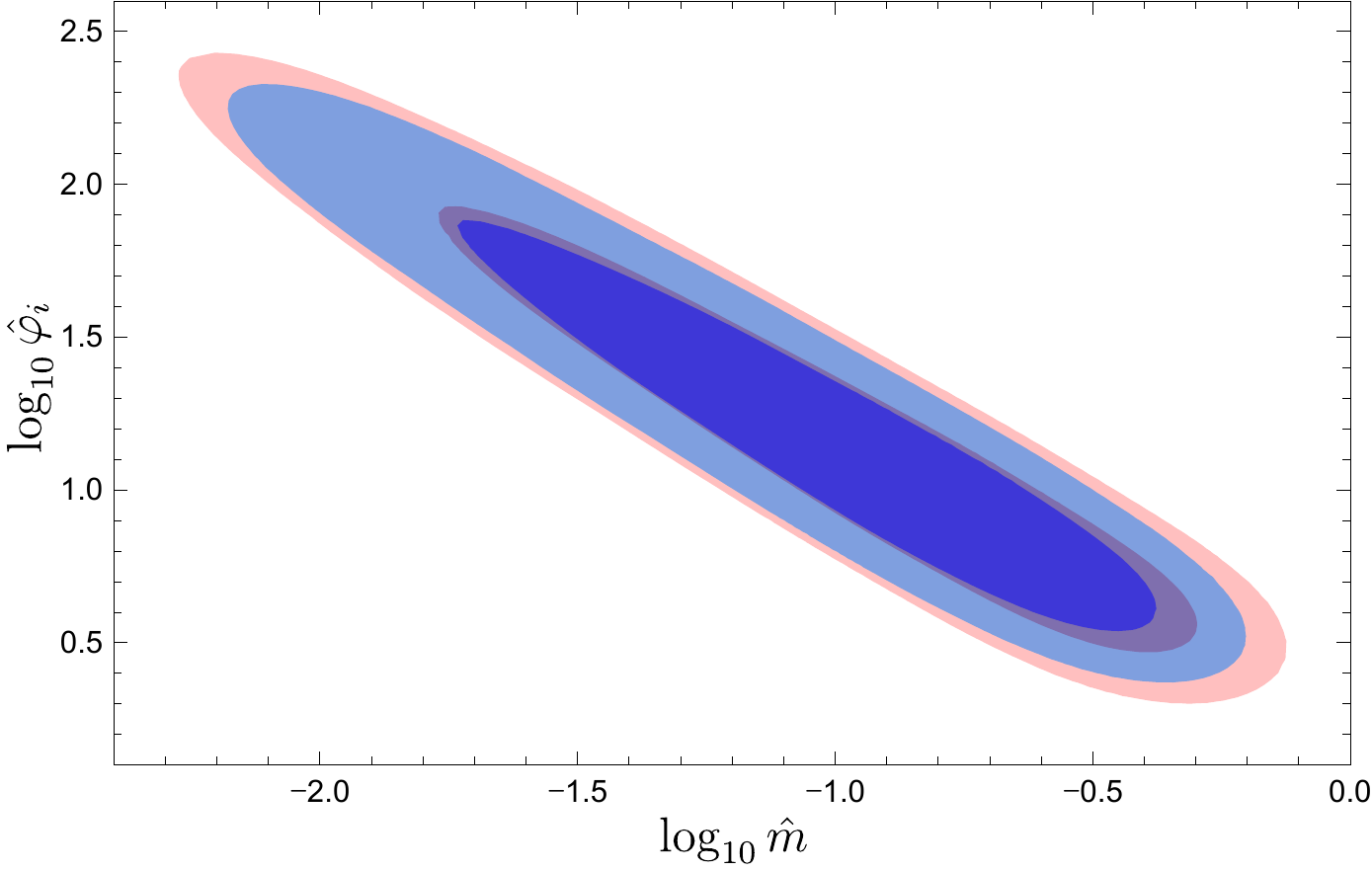}}
\end{center}
\caption{\small
Marginalized likelihood distributions for $(H_0, \log_{10} \hat{m})$ (left) and $(\log_{10} \hat{\varphi}_i, \log_{10} \hat{m})$ (right) with 68.3\% and 95.4\% confidence limits, obtained by the joint parameter estimation with $H(z)$ + SN+BAO+PLANCK (blue) and $H(z)$+SN+BAO+WMAP9 (red) data sets, respectively.
}
\label{fig:like2d}
\end{figure}

We explore the allowed ranges of our dark energy model parameters using the recent observational data by applying the MCMC parameter estimation method. In the calculation, we use $\log_{10}\hat{m}$, $-\log_{10}(-\chi)$, $\Omega_m h^2$, $\Omega_b h^2$, and $\log_{10}\hat{\varphi}_i$ as free parameters. The results are shown in Table \ref{table:result1}
for a summary of parameter constraints with mean and $1\sigma$ confidence limits and in Fig.\ \ref{fig:like} for marginalized one-dimensional likelihood distributions of individual parameters. We can see that the result obtained with Planck data
gives tighter constraints on model parameters.
The best-fit locations in the parameter space are
\begin{align}
(\log_{10}\hat{m}, -\log_{10}(-\chi), \Omega_m h^2, \Omega_b h^2, \log_{10}\hat{\varphi}_i)
=(-1.314, 6.964, 0.141, 0.024, 1.47) \,,
\end{align}
with a minimum chi-square of $\chi_\textrm{min}^2 = 588.391$ for the $H(z)$+SN+BAO+WMAP9,
and
\begin{align}
(\log_{10}\hat{m}, -\log_{10}(-\chi), \Omega_m h^2, \Omega_b h^2, \log_{10}\hat{\varphi}_i)
=(-1.270, 6.908, 0.145, 0.024, 1.42) \,,
\end{align}
with $\chi_\textrm{min}^2 = 590.804$ for $H(z)$+SN+BAO+PLANCK. The behaviors of Hubble parameter and SN distance modulus as a function of redshift are shown in Fig.\ \ref{fig:bestH}.
In Fig.~\ref{fig:like2d}, we also present the marginalized likelihood distributions for $(H_0, \log_{10} \hat{m})$ and $(\log_{10}\hat{\varphi}_i,\log_{10} \hat{m})$, which shows that the value of Hubble constant does not depend on the variation of photon mass while the initial value of $\hat{\varphi}$ decreases as the photon mass increases.

To assess the goodness-of-fit of our massive photon model, in Table \ref{table:result1} we present the parameter constraints for the $\Lambda\textrm{CDM}$ model and list the value of the minimum reduced chi-square ($\chi_\nu^2$) for each case. The minimum reduced chi-square is defined as $\chi_\nu^2=\chi_{\textrm{min}}^2 /\nu$, where $\nu=N-n-1$ is the number of degrees of freedom and $N$ and $n$ are the numbers of data points and free model parameters, respectively. In our analysis, $N=618$, and $n=5$ for our massive photon model and $n=3$ for the $\Lambda\textrm{CDM}$ model. Although the simple $\Lambda\textrm{CDM}$ model gives the slightly better fit to the observational data with the smaller values of $\chi_\textrm{min}^2$ and $\chi_\nu^2$, we judge that our massive photon model fits the data reasonably well in the sense that the reduced chi-square is very close to unity.


We note that for our model to be compatible with observations the photon should have non-zero mass with $\log_{10}\hat{m}\approx -1$, which corresponds to the photon mass $m \approx 10^{-34}$ eV.
Such a value is consistent with current experimental upper bound on the photon mass
$m \leq 10^{-15}$~eV from the measurements of Earth's magnetic field~\cite{Fischbach:1994ir},
Pioneer-10 data of the Jupiter magnetic field~\cite{Davis:1975mn}, 
and $m \leq 10^{-27}$~eV from the galactic magnetic fields~\cite{GalMagField}.

\begin{table*}
\begin{center}
\caption{Summary of parameter constraints }
\label{table:result1}
\resizebox{\columnwidth}{!}{%
\begin{tabular}{c||cc|cc}
\hline\hline 
     &   \multicolumn{2}{|c|}{Massive Photon Model}    &   \multicolumn{2}{|c}{$\Lambda{\rm CDM}$ Model}   \\
\hline
 & $H(z)$ + SN + BAO &  $H(z)$ + SN + BAO & $H(z)$ + SN + BAO & $H(z)$ + SN + BAO \\ 
 & + WMAP9 & + PLANCK & +WMAP9 & +PLANCK\\
\hline 
$H_0$ & $69.57^{+0.84}_{-0.85}$ &     $69.21^{+0.71}_{-0.66}$   & $69.57^{+0.83}_{-0.80}$     & $69.32^{+0.67}_{-0.69}$\\

$\log_{10}\hat{m}$  & $-0.8089^{+0.3718}_{-0.5758}$ & $-0.8400^{+0.3200}_{-0.5393}$  & - & - \\

$-\log_{10}(-\chi)$ & $> 4.1~(2 \sigma)$  & $> 4.3~(2 \sigma)$ & - & -\\

$\Omega_m h^2$ & $0.1409^{+0.0024}_{-0.0024}$ & $0.1448^{+0.0016}_{-0.0014}$ & $0.1410^{+0.0022}_{-0.0024}$& $0.1446^{+0.0014}_{-0.0015}$\\

$\Omega_b h^2$  & $0.0240^{+0.0005}_{-0.0004}$ &  $0.0239^{+0.0003}_{-0.0003}$ & $0.0239^{+0.0004}_{-0.0004}$& $0.0239^{+0.0003}_{-0.0003}$\\

$\log_{10}\hat{\varphi}_i$  & $0.9586^{+0.5751}_{-0.3620}$ &   $0.9898^{+0.5364}_{-0.3161}$ & - & -\\

$\Omega_\Lambda h^2$  
 & - & - 
 & $0.3433^{+0.0120}_{-0.0118}$ & $0.3355^{+0.0105}_{-0.0105}$ \\

\hline $\chi^{2}_\textrm{min}$  & 588.391 &    590.804  & 588.366 & 590.724\\
\hline $\chi^{2}_\nu$  & 0.96142 &   0.96536  & 0.95825 & 0.96209\\
\hline
\hline
\end{tabular}
}
\end{center}
\end{table*}

\vspace{-.0cm}

\section{Conclusion}\label{seccon}

In this paper, we investigated the cosmological implications of the massive Stueckelberg QED non-minimally coupled to the Einstein gravity, paying a special attention to the possible role of massive photon in  relation with the dark energy. We found that the theory allows a long period of current accelerating phase which closely mimics $\Lambda$CDM in which the acceleration of Universe is due to the nonvanishing photon mass governed by the relation $\Lambda\sim m^2$. A detailed numerical analysis comparing with the various data predicts the nonvanishing photon mass being of the order of $\sim 10^{-34} {\rm eV}$, which is consistent with the other upper limits available so far.

The cosmological evolution of the non-minimal SVT gravity theory exhibits a couple of interesting properties. The first of Fig. 2 shows that the dark energy density of  \eqref{energypressure} has the same scaling behavior with the radiation energy density in the radiation-dominated epoch.
Also during the  intermediate state between radiation and constant dark energy epoch,
the behavior of dark energy density mimics the pressureless matter
which can be seen clearly from the equation of state graph of Fig.~\ref{fig:energyEOS}.
 Then, this constant dark energy dominant era lasts for a long period of time (Fig.~\ref{fig:energyEOS}),
in which the current acceleration of Universe takes place.
During this period, the dark energy density is practically given by an intriguing relation $\rho^{\rm (de)}\sim m^2M_p^2.$

We note that 
the scalar field stays almost constant (Fig.~\ref{fig:fphisol}) before a relaxation  to its natural value 0 begins to occur
during the matter-dominated epoch.
Analysis in Sec.~\ref{anaDE} shows that
the Stueckelberg scalar field will ultimately relax to zero after going through a period of oscillations.
Both the energy density and pressure decays as $1/a^3$ during the oscillations,
but the pressure (and the temporal component) also oscillates in harmony with the scalar fields. Therefore, the analysis predicts a gradual deviation from $\Lambda$CDM in the future and the Universe will see the return of matter-dominated
epoch (not the pressureless dust but the remnants of the scalar-temporal field component
oscillations).

We also compared the massive photon model with the observational data of
SN Ia, Hubble parameter, BAO and CMB measurements.
According to MCMC methods, we obtained the best fit values of the parameters~(shown in Table~\ref{table:result1}) by
fixing the value of $\hat{\eta}$ to $0.9$ and $\hat{\omega}$ to $-0.35$.
It may be important to mention here that this fixed values of parameters $\hat{\eta}$ and $\hat{\omega}$
correspond to the decay mode during the matter dominated epoch.
Presumably, different values will not alter the numerical results
much as long as
these parameters are chosen to satisfy the decay condition~\eqref{condbeta}, $\beta<3$.
We found that $m\sim 10^{-34} {\rm eV}$ is allowed by
the $H(z)$ + SN + BAO + CMB dataset
for the massive Stueckelberg QED non-minimally coupled to the Einstein gravity.
This is consistent with the most stringent upper bounds on the photon mass listed by
the Particle Data Group~\cite{Hagiwara:2002fs}.
In addition, this result can give a high precise estimation for the mass of the photon.
We also found that the  $\Lambda$CDM model is still compatible
with our massive photon model.

We conclude with a final comment on $\rho^{{\rm (de)}}\vert_0 \sim \Lambda
M_p^2\sim m^2M_p^2$.
It would be certainly impossible to perform  any experiment which
establishes the exact vanishing of the photon mass, but the  ultimate upper limit on the photon rest mass, $m$, can be estimated by using the uncertainty principle to be $m \approx \hbar/(\Delta t) c^2\cong 10^{-34}$ eV for the current age of the universe. Our analysis with the observational data shows that this value is in agreement with the prediction of massive QED. It is also interesting to note that the relation $\Lambda\sim m^2$
provides a vacuum energy density $\Lambda_{c}^4\sim\Lambda M_{\rm p}^2$ with IR cutoff $L\sim m^{-1}$ in accordance with the holographic
constraint \cite{Cohen:1998zx}.

\section{Acknowledgments}
We like to thank the anonymous referee for valuable comments and C. Lee and W.-T. Kim for useful discussions. 
P.O. was supported by Basic Science Research Program through the National Research Foundation of Korea (NRF) funded by 
the Ministry of Education(2015R1D1A1A01056572).
C.G.P. was supported by Basic Science Research Program through the National Research Foundation (NRF) of Korea funded 
by the Ministry of Science, ICT and Future Planning (No.\ 2013R1A1A1011107).


\end{document}